\documentclass{JINST}

\usepackage{units}		
\usepackage{comment}	

\title{Spectral Characterization and Modeling of Wavelength-shifting Fibers}

\author{	R.~B.~Pahlka$^{a,b}$, 
		G.~Elpers$^a$,
		J.~Huang$^{a,c}$,
		K. Lang$^a$,\thanks{Corresponding author.}~ 
		M. Proga$^a$\\
\llap{$^a$}University of Texas at Austin,
		 Department of Physics C1600,
		 1 University Station, Austin, TX 78712-0264, USA\\
\llap{$^b$} now at Texas Children's Hospital, Houston, TX 77030, USA\\
\llap{$^c$} now at Universt{\"a}t Z{\"u}rich, Z{\"u}rich, Switzerland\\
  E-mail: \email{lang@physics.utexas.edu}}

\abstract
{
 We have constructed a detailed software model of photon transport in clear and wavelength shifting polystyrene optical fibers based on the GEANT4 framework.  We validate the model with measurements obtained from several different lengths of fiber illuminated with several LEDs, using two independent acquisition systems.  The simulated spectral and spatial light distributions at the end of the fiber and the resultant attenuation lengths show good agreement with measurement.  We use the model to predict the fiber behaviour for several test cases and discuss potential applications.   
}

\keywords{scintillation; fibers; wavelength-shifting; photomultiplier; plastic scintillator; optical photon transport; GEANT4; double beta decay}

\begin{document}


\section{Introduction}

Optical wavelength shifting fibers are now common components of liquid scintillator- and extrusion-based detector technologies in high energy physics experiments.  Issues of dopant material and concentration have been addressed~\cite{PlaDalmau:1994kx}, and light attenuation properties of plastic optical fibers have been discussed both in experimental~\cite{Santos,Adamian:2004bz,Ranucci:2002vs} and theoretical~\cite{Ranucci_2,Cavasinni:2004rr} frameworks.  The properties of common fiber materials including polystyrene~\cite{Kabanov,D'Ambrosio:1991iy,Kaino}, polymethylmethacrylate (PMMA)~\cite{Polishuk,Swalen}, and fluorinated polymer~\cite{Jiurong} are well known.  Most mechanisms and processes with respect to energy transfer~\cite{Hirayama}, Rayleigh scattering~\cite{Bunge}, Stokes shifting~\cite{Lakowicz,Nikitina}, optical dispersion~\cite{Kasarova:2007,Cochrane}, and molecular orientation~\cite{Llop} have been elucidated.  

Several models of photon transport and fiber simulation have been introduced in various frameworks including dynamic programming~\cite{Calvo}, Monte Carlo~\cite{Argyriades:2010wd,Cavasinni:2004rr,Tickner:2007}, finite element analysis~\cite{Koeppen}, and sensitivity analysis~\cite{Ghal-Eh,Frlez:2000aw}.  Others have considered simulations of electronic readout systems~\cite{Sanchez:2010zz} and photomultiplier tubes~\cite{Creusot:2010xe}.  Moreover, the literature on multimode optical fiber for use in data transmission is vast~\cite{Kutz}.

We are motivated by the need to accurately predict the spectral properties of light exiting the fiber to ensure that these spectra can be matched to the appropriate photodetector.  Given the inherent large-scale nature of current and future detectors employing long WLS fibers and the notion that many thousands of meters of fiber are used in the construction, this work serves as an important aid to the overall design, particularly for next generation detectors.  

Moreover, a condition of the employment of wavelength shifting fibers in high energy physics experiments is the requirement of fast and efficient fluorophores in the region of high optical transparency.  Studies of new wavelength shifters has met with some success~\cite{PlaDalmau:1994kx} but continues to remain challenging~\cite{Kaufmann:1997gm}.  The flexibility of this model allows the study of the spectral distribution of light after propagation, and as seen by the photodetector, for any supplied fluorophore emission spectrum.  This is particularly useful in studying other fluorescent compounds emitting at long wavelengths (yellow or red) as well as the green-emitting family of coumarins~\cite{Trenor}. 

In this work, we introduce a general model of photon transport based on the GEANT4 framework~\cite{Agostinelli:2002hh} that accounts for all of the processes mentioned above for clear and wavelength shifting plastic optical fiber.  We perform a preliminary validation of the model using measurements from Kuraray clear fibers illuminated with several LEDs to confirm spectral output and attenuation agreement with simulation.  We further validate the model using measurements from Kuraray Y-11 wavelength shifting fibers of various lengths then use the model to predict the spectral output behaviour from fibers with various concentrations, diameters, and bending radii.  We demonstrate that by including the spectral properties of all components and properly accounting for absorption, reemission, and fluorescent quantum yield of the chromophore, this model can accurately determine the spectral response, timing, and effective attenuation in wavelength shifting fibers.

\section{Materials and methods}

\subsection{Fiber preparation}

We used clear fibers and Y-11 Non-S type wavelength shifting fibers from Kuraray~\cite{Kuraray_1} of several diameters, lengths, and dopant concentrations in these studies.  Both types of fibers are dual clad.  The core is composed of polystyrene with a refractive index of 1.59 at 500\,nm and a density of 1.05\,g/cm$^3$.  The inner cladding is composed of poly-methylmethacrylate (PMMA) with a refractive index of 1.49 at 500\,nm and a density of 1.19\,g/cm$^3$.  The outer cladding is composed of a PMMA-based fluorinated polymer with a refractive index of 1.42 at 500\,nm and a density of 1.43\,g/cm$^3$.  Each cladding thickness is 3\% of the diameter.  

The ends of each fiber were polished using a diamond-tipped cutter to obtain uniform light collection~\cite{Chardon}.  A small delrin collet was attached to each end for accurate positioning with respect to the readout detectors employed and the light source.  We illuminate the fiber with light oriented perpendicularly to the principal fiber axis by securing one end of a fiber in a plastic block into which a trench of appropriate size has been drilled. An LED was placed in an adjacent block, oriented perpendicular to the fiber axis for illumination. 

\subsection{LED characterization}

\begin{figure}[h]
\centerline{
\includegraphics[width=0.45\linewidth]{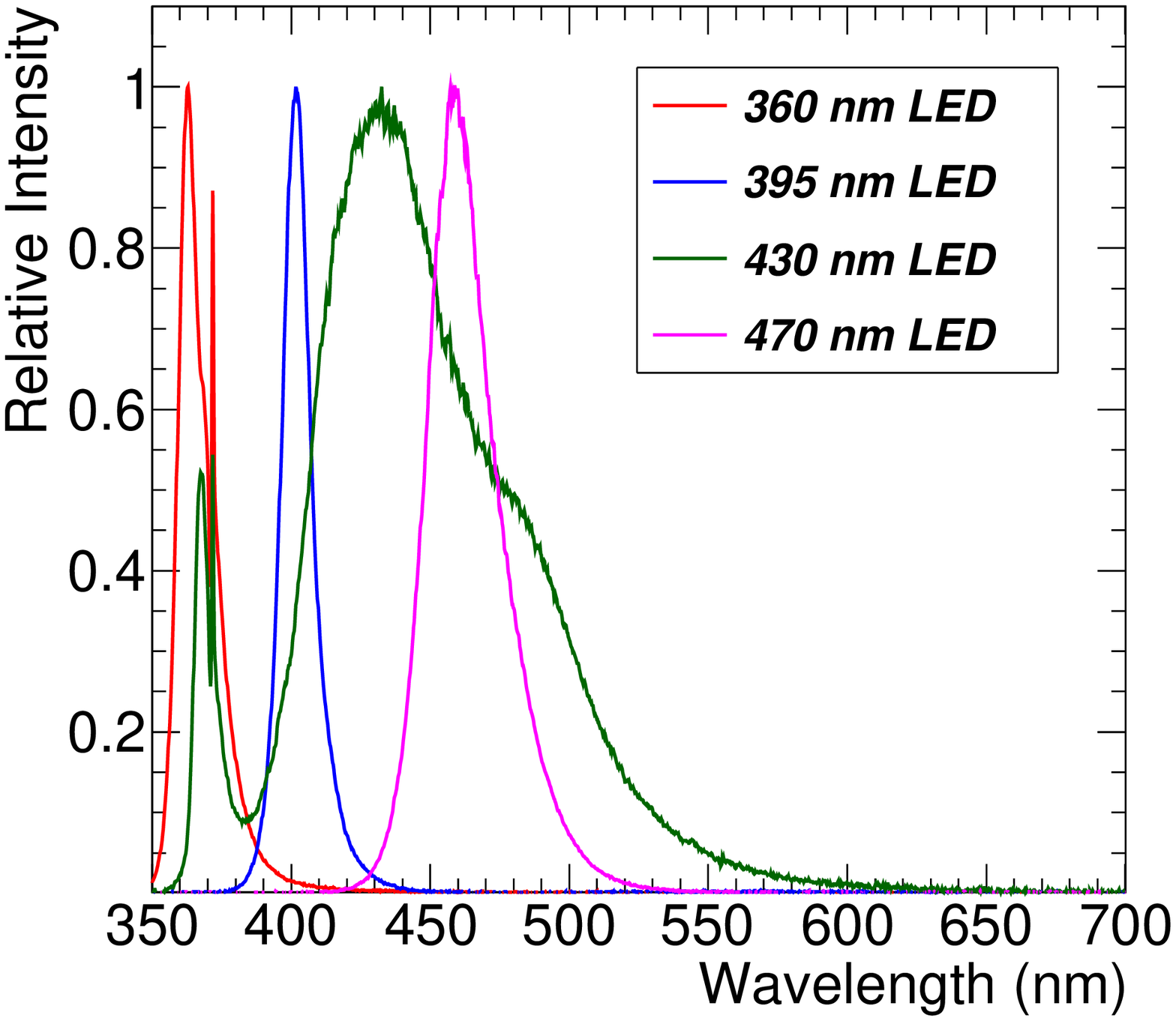}
\hskip0.25in
\includegraphics[width=0.45\linewidth]{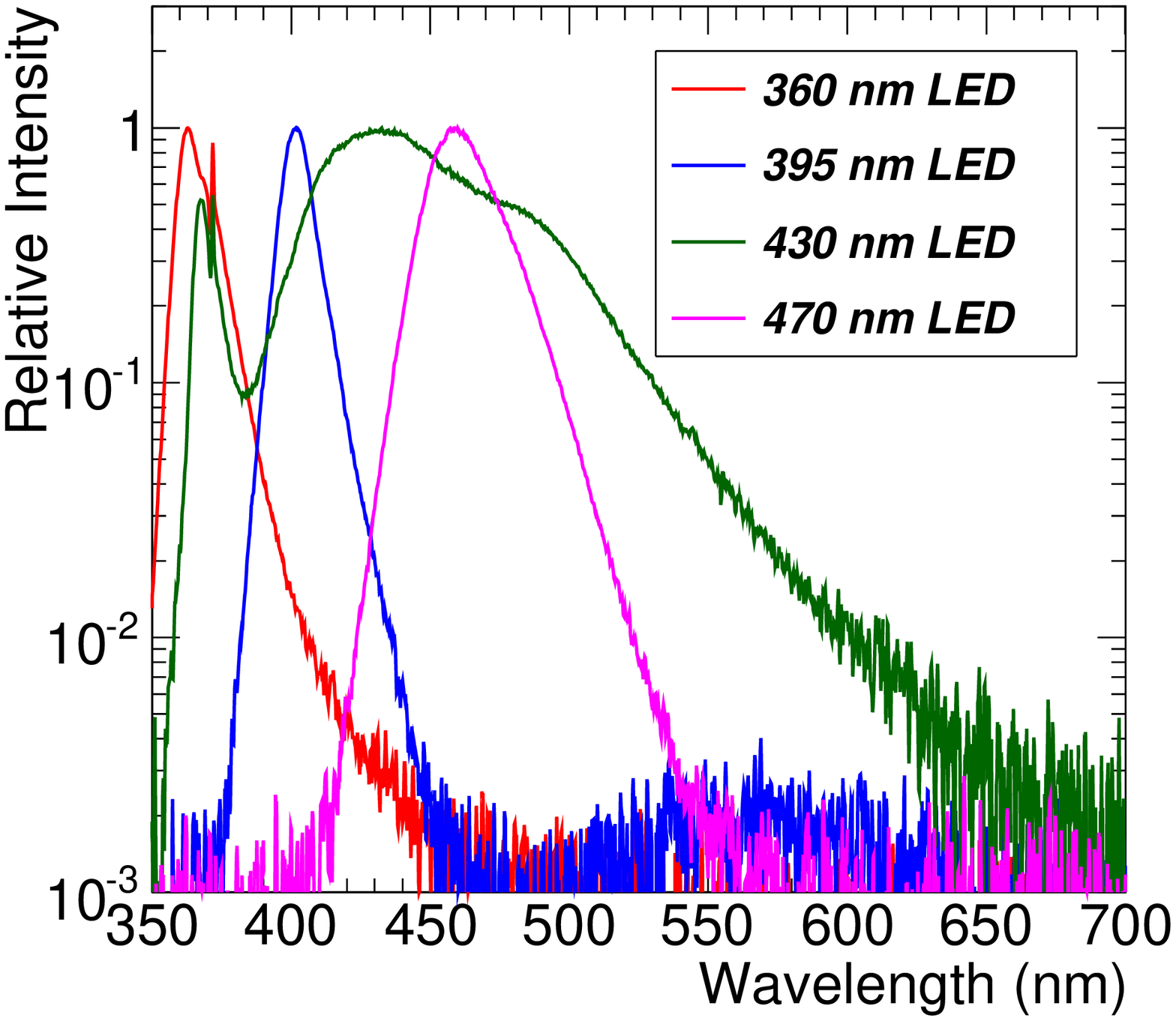}
}
\caption{Emission spectra for four LEDs as measured by an Ocean Optics USB-4000 spectrophotometer.  
The spikes at 370\,nm are inherent features of both the 360\,nm and 430\,nm LEDs.  All LEDs exhibit significant emission in the tails. }
\label{fig:led_spectra}
\end{figure}

We used several different light emitting diodes (LEDs) from several manufacturers [REF] to illuminate the fibers and obtain the spectral response.  The LEDs were pulsed with short square wave pulses using a LeCroy 9210 pulse generator with a LeCroy 9212 300\,MHz, 300\,ps variable edge output module.  All LED emission was optimized for frequency, applied voltage, and pulse width.  LEDs were illuminated and aimed at the spectrometer at a distance of approximately 12~inches through air.  Spectra for the LEDs used for fiber illumination are shown in Figure~\ref{fig:led_spectra} with all curves normalized to peak.  All LEDs exhibit spectral widths ranging from 20 to 70\,nm, some with very broad tails.   We chose the 430\,nm LED as a reference since it closely matches many of the primary fluorescence spectra of standard plastic scintillators. The LED spectra used as inputs in the simulation were obtained using the output spectra obtained from the spectrophotometer.  

\begin{figure}[h]
\centerline
{
\includegraphics[width=0.7\linewidth]{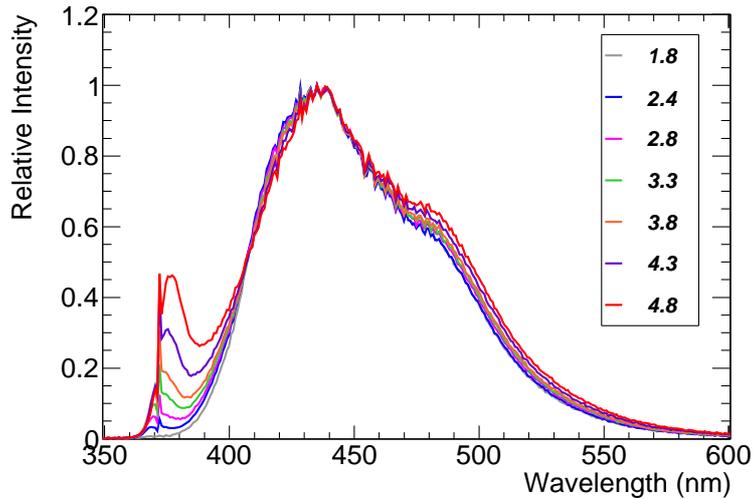}
}
\caption{Emission spectra for the 430\,nm LED measured by an Ocean Optics USB-4000 
spectrophotometer for several applied voltages. All spectra are normalized to the peak emission. }
\label{fig:led_voltages}
\end{figure}

Figure~\ref{fig:led_voltages} shows the emission spectra for the 430\,nm LED measured by the spectrophotometer for several applied voltages.  The peak wavelength is the same in all cases, however higher applied voltages enhance the short wavelength region of the emissions spectrum and tend to suppress emission in the long wavelength region.  

\subsection{Spectrophotometer and spectrometer data acquisition}

We measure the spectral response from fibers using the Ocean Optics USB-4000 spectrophotometer. The spectral response of this detector is approximately 370--1100\,nm with a resolution of 0.3\,nm (FWHM).  Acquisition was performed using the SpectraSuite software from Ocean Optics.  
We also measured the spatial and spectral distributions of light exiting fibers using a Spec10:100 BR-N CCD camera from Princeton Instruments.  Each pixel size is 20\,$\mu$m$\times$20\,$\mu$m with 1340 columns and 100 rows and the spectra obtained have been corrected for quantum efficiency.  We couple the CCD camera to a Horiba Jobin-Yvon iHR550 imaging specrometer.  We measure the spatial and spectral distributions from 0.7\,mm 1 meter 200\,ppm Y-11 WLS fibers.  One end of the fiber is placed at the entrance slit to the spectrometer.  Near the other end, we illuminate the fiber on the side with a 430\,nm LED.

\section{Spectral dependence of absorption and emission}

\subsection{Absorption spectra}

Several different loss mechanisms contribute to the overall absorption curve in clear polystyrene fibers, which is almost entirely determined by the properties of the core material~\cite{Emslie}.   Figure~\ref{fig:emslie_plot} shows a representation of these mechanisms as a function of wavelength.  The shapes arise predominantly from wavelength-independent scattering, Rayleigh scattering, and ultraviolet and infrared absorption. A thorough discussion is presented in~\cite{Emslie}.  Wavelength-independent scattering includes absorption effects due to dust, cracks, and bubbles, and the fiber transparency generally depends on the manufacturing process.  In this case, the diameters of the inclusions issuing the effect are on the order of one micrometer or larger.  This is contrasted with Rayleigh scattering which shows a pronounced wavelength dependence caused by smaller irregularities on the order of one-tenth of a wavelength.  Loss due to Rayleigh scattering is inversely proportional to the fourth power of the wavelength~\cite{Emslie,Kaino}.  The effect of Rayleigh scattering (and Mie scattering) on the spatial and angular distributions of light from polymer optical fiber have been described with theoretical models and compared with measurement~\cite{Bunge}.  Ultraviolet absorption arises from electronic transitions between energy levels in benzene rings~\cite{Emslie} and the loss is typically described empirically~\cite{Kaino}.  Finally, infrared absorption is due to both aliphatic and aromatic vibrational harmonics in the polystyrene carbon-hydrogen bonds~\cite{Emslie,Kaino}. 

\subsection{Fluorescence emission spectra}

While the manufacturing process is complex, wavelength-shifting fibers are realized by simply doping the clear polystyrene with an appropriate fluorescent molecule. We consider standard fluorescent principles where a) the emission spectrum is independent of the excitation wavelength, b) there exist non-zero and finite energy losses between excitation and emission, and c) the absorption and emission spectra are typically mirror-images with the absorption (and hence emission) peaks are due to the vibrational energy level spacing within the molecule.  Here, transitions from the lowest vibrational level of the ground state to the higher vibrational levels of the first excited state engender the peaks~\cite{Lakowicz}.  

\begin{figure}[h]
\centerline
{
\includegraphics[width=0.45\linewidth]{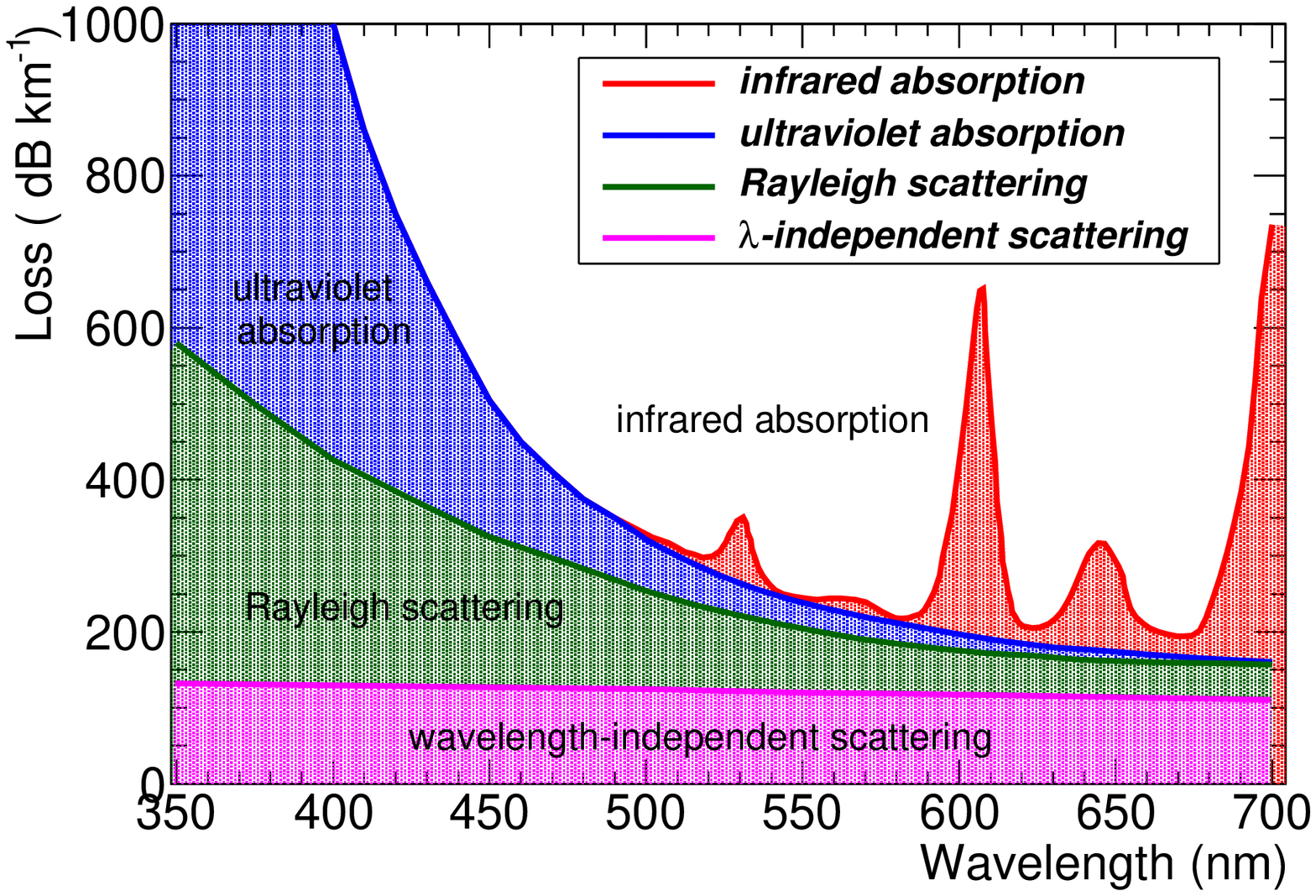}
\hskip0.25in
\includegraphics[width=0.45\linewidth]{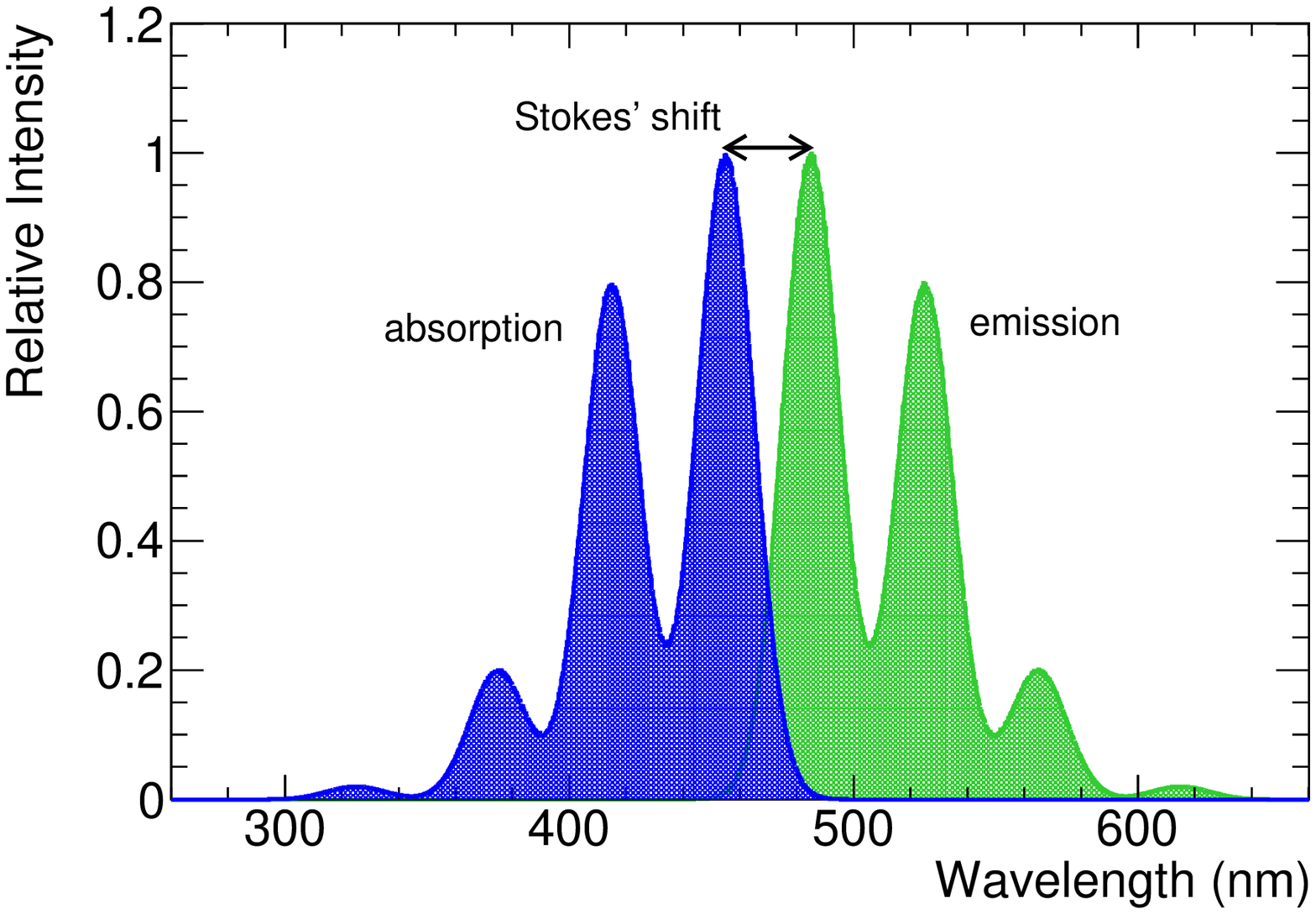}
}
\caption{Left: Absorption curves for several processes contributing to the overall absorption length for clear polystyrene fibers.  Components include wavelength-independent scattering, Rayleigh scattering, ultraviolet absorption and infrared absorption bands.  This representation has been generated similar to what is presented in~\cite{Emslie} and only serves as an approximation to the actual absorption length of Kuraray clear fibers. Right: Idealized absorption and emission spectra for an arbitrary fluorescent molecule showing mirror symmetry and a characteristic Stokes' shift. See for example, treatments in~\cite{Wright,Angulo}. }
\label{fig:emslie_plot}
\end{figure}

For fluorescence, the energy yield is always less than unity because of Stokes' losses and can be quantified by the fluorescent quantum yield $Q$, which is simply the ratio of the radiative decay rate to the total rate:
$$
Q = \frac{k_{r}}{k_{r} + k_{nr}}
$$
\noindent where $k_{nr}$ is the collective non-radiative decay processes and $k_{r}$ is given by the Strickler-Berg equation for intrinsic radiative decay~\cite{Strickler,Lakowicz}:
$$
k_{r} = \frac{1}{\tau_{r}^{0}} = \frac{8\pi 230 c n^2}{N_{A}}\frac{\int F(\nu) d\nu}{\int(F(\nu)/\nu^{3})d\nu}\int\frac{\epsilon(\nu)}{\nu}d\nu
$$
\noindent $F$ is the fluorescence spectrum corrected for instrument response, $\epsilon$ is the absorption spectrum, and $n$ is the refractive index. This formalism works well modulo a few caveats~\cite{Lakowicz}.  While there are also exceptions to the mirror-image rule, Figure~\ref{fig:emslie_plot} also shows idealized absorption and emission spectra for an arbitrary fluorescent molecule. 


The fluorescent absorption and emission distributions of generic fluorescent molecules can be calculated from first principles using a variety of models.  One in particular, has been illustrated previously using a model whose molecular vibrational wavefunctions are based on a displaced-distorted harmonic oscillator, convolved with a Gaussian line broadening function~\cite{Angulo}.  The ground state vibrational wavefunction with $a$ proportional to the displacement and $R$ proportional to the distortion is given by:
$$
f(x) =  \left ( \frac{R}{\pi} \right )^{1/4} e^{\left(-\frac{(x+a)^{2}R}{2} \right )}
$$
The excited vibrational state basis functions are taken as Hermite polynomials each with a normalization factor $N_{\nu^{'}}$:
$$
\psi_{\nu}(x) = N e^{-(x^{2}/2)} \left [(-1)^{\nu} e^{x^2} \frac{d^{\nu}}{dx^{\nu}} e^{-x^2} \right]
$$
\noindent with the normalization factor $N$ given by
$$
N_{\nu^{'}} = \frac{1}{\sqrt{2^{\nu^{'}}}\nu^{'}\pi^{1/2}}
$$
\noindent The factors $c_{0\nu^{'}}$ are the Franck-Condon factors describing the projection of the ground state onto the excited state and are calculated as:
$$
c_{0\nu^{'}} = \int_{-\infty}^{\infty} f(x)\psi_{\nu^{'}}(x)dx
$$
\noindent This describes the overlap between the two vibrational states which dictates the absorption amplitude~\cite{Wright}.  Summing over all possible states gives the absorption (and emission) transition distributions from the ground vibrational state ($\nu^{''}$) to excited vibrational states ($\nu^{'}$) (and vice-versa) and are given by:
$$
A(\nu^{''} = 0 \rightarrow \nu^{'}) = N \sum_{\nu^{'} = 0}^{\infty} c_{0\nu^{'}}^{2}\sqrt{\frac{2}{\pi \omega^{2}}} exp(-\frac{2(\bar{\nu} - \bar{\nu}_{abs(flr)}^{0-0} - \nu^{'}\bar{\nu_{vib}})^{2}}{\omega^{2}}).
$$
\noindent where $\omega$ is the Gaussian width, $\bar{\nu}_{abs}^{0-0}$ is the 0-0 absorption wavenumber, $\nu_{vib}$ is the constant frequency wavenumber and $\bar{\nu}$ is the spectral wavenumber.   The equation above can be used to calculate either absorption (abs) or emission (flr) spectra.  Finally, the Lippert equation is introduced~\cite{Lakowicz} to account for potential solvent effects.  This is given as:
$$
\bar{\nu}_{abs(flr)}^{0-0} = \bar{\nu}_{g}^{0-0} - \frac{2}{4\pi \epsilon_{0}hc}\frac{1}{a^{3}}\mu_{g(e)}(\mu_{e} - \mu_{g})F(n^{2},\epsilon) 
$$
\noindent with $\bar{\nu}_{g}^{0-0}$ being the 0-0 transition in absence of solvent, $a$ the cavity radius in which the solute is embedded, $\mu_{g}$ and $\mu_{e}$ the modules of the dipole moments of the molecules in the ground and excited states, respectively, and $F(n^{2},\epsilon)$ for the solvent function.   The above treatment shows one example of how fluorescence absorption and emission spectra can be calculated and, with the introduction of the Lippert equation, shows how the solvent engenders potential systematic shifts in the emission spectra, which is discussed later in the context of the model.  


\section{Description of the model}

We incorporate the spectral properties of all materials into the simulation and use the GEANT4 framework as the foundation of the model.  In this framework, two attenuation lengths must be specified: one for processes where the primary photon is absorbed in polystyrene, and one for processes where the primary photon is absorbed by the wavelength shifting fluorescent compound and re-emitted.  The absorption/re-emission process can occur multiple times. The WLS fluorescent compound used is K27 which has a fluorescent quantum yield of 0.7~\cite{PlaDalmau:1994kx}.   We account for the fluorescent quantum yield of K27 by allowing absorbed photons to be re-emitted with 70\% probability~\cite{PlaDalmau:1994kx} and at a wavelength equal to or greater than the absorbing wavelength for energy conservation.  Polystyrene exhibits a small amount of fluorescence which we neglect in this model~\cite{Kabanov,Hirayama}.  

\begin{figure}[h]
\centerline
{
\includegraphics[width=0.45\linewidth]{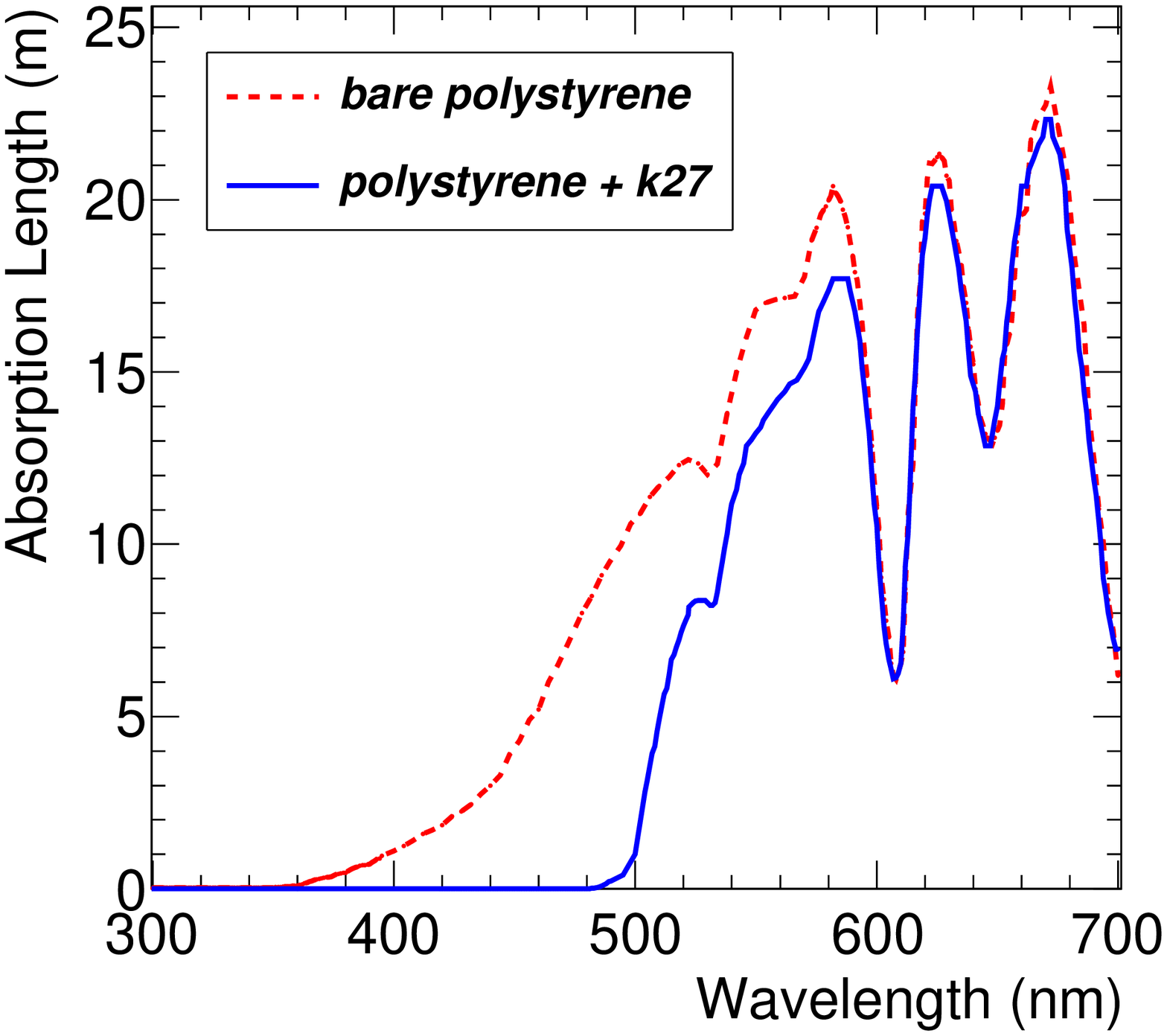}
\hskip0.25in
\includegraphics[width=0.45\linewidth]{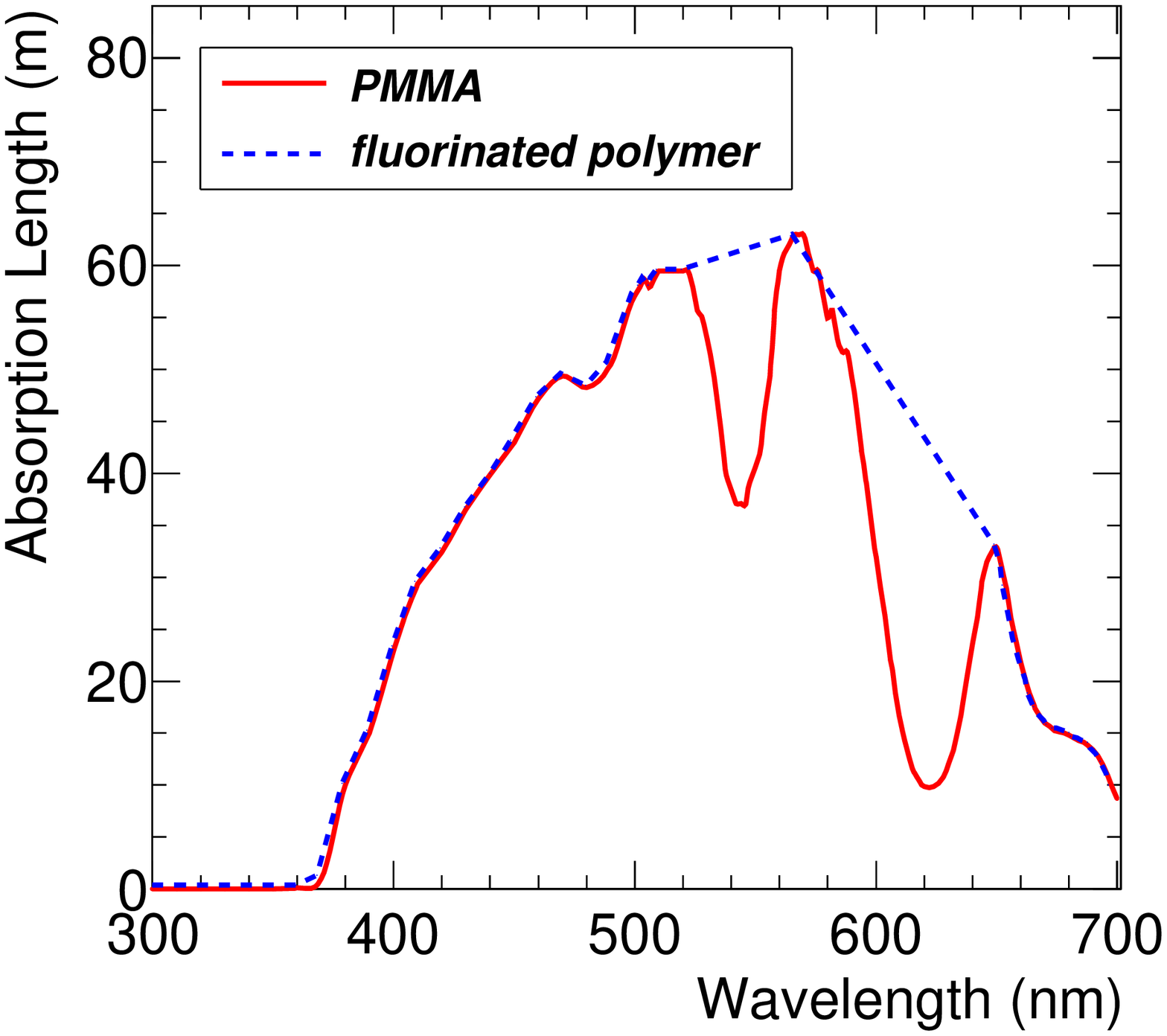}
}
\caption{Left: Absorption lengths for K27-doped and un-doped polystyrene estimated from Kuraray data~\cite{Kuraray_2}. 
Right: Absorption lengths for PMMA and fluorinated polymer.  
The absorption bands in each case are attributed to various carbon-hydrogen bonds in the polymers.  }
\label{figure:polystyrene}
\end{figure}

Figure~\ref{figure:polystyrene} shows the absorption spectra of clear polystyrene, K27-doped polystyrene (at 200 ppm), PMMA, and fluorinated polymer.  The absorption spectra of the doped and un-doped polystyrene was taken from Kuraray~\cite{Kuraray_1}.   The absorption spectrum of PMMA was taken from~\cite{Polishuk} although it has been measured elsewhere~\cite{Koeppen}.  The absorption spectrum of the fluorinated polymer has been estimated to be that of PMMA with the condition that the absorption bands due to the carbon-carbon bonding has been removed.  Surface fluorination has been studied previously~\cite{Jiurong,Koeppen}.

\begin{figure}[h]
\centerline{
\includegraphics[width=0.45\linewidth]{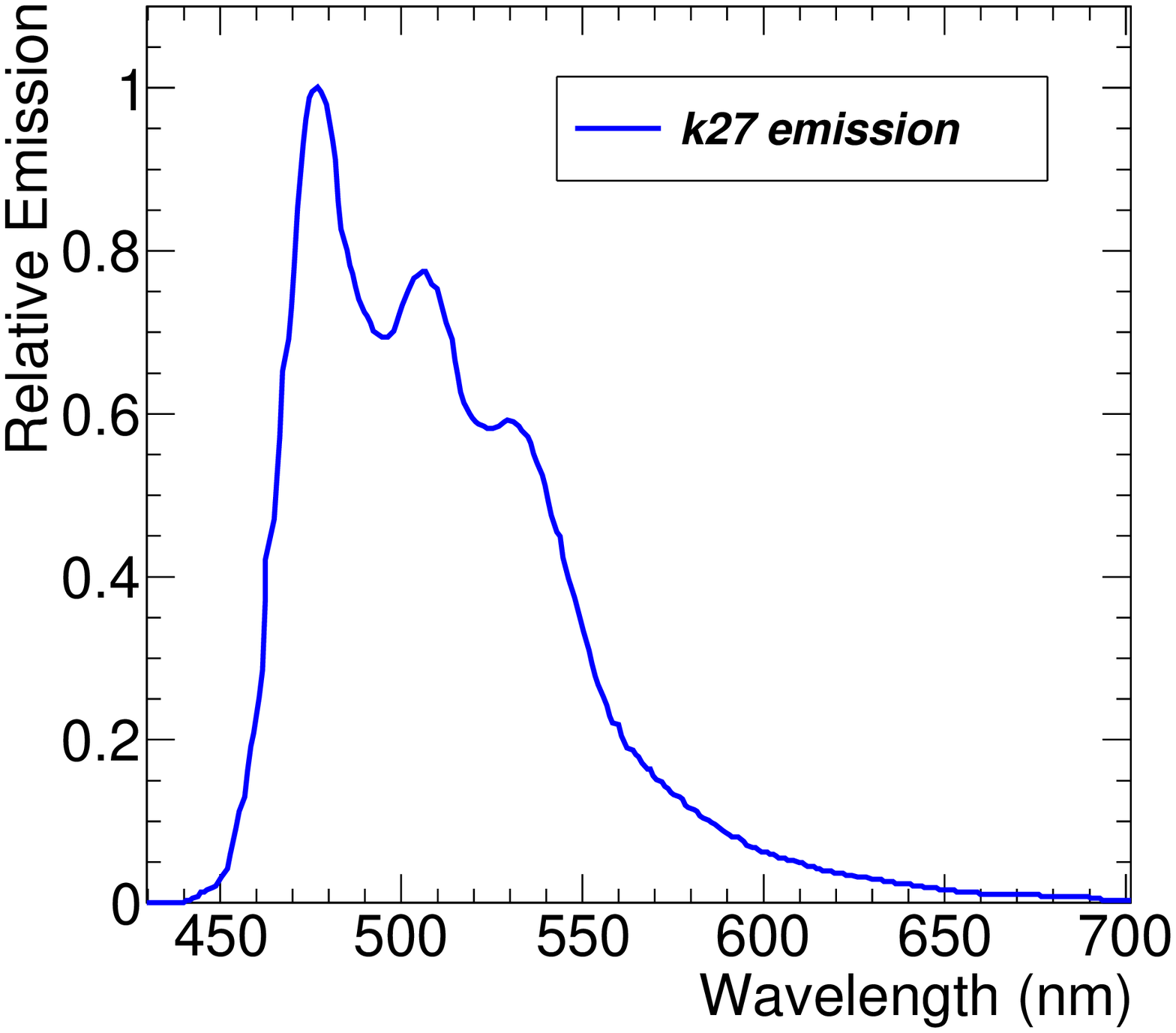}
\hskip0.25in
\includegraphics[width=0.45\linewidth]{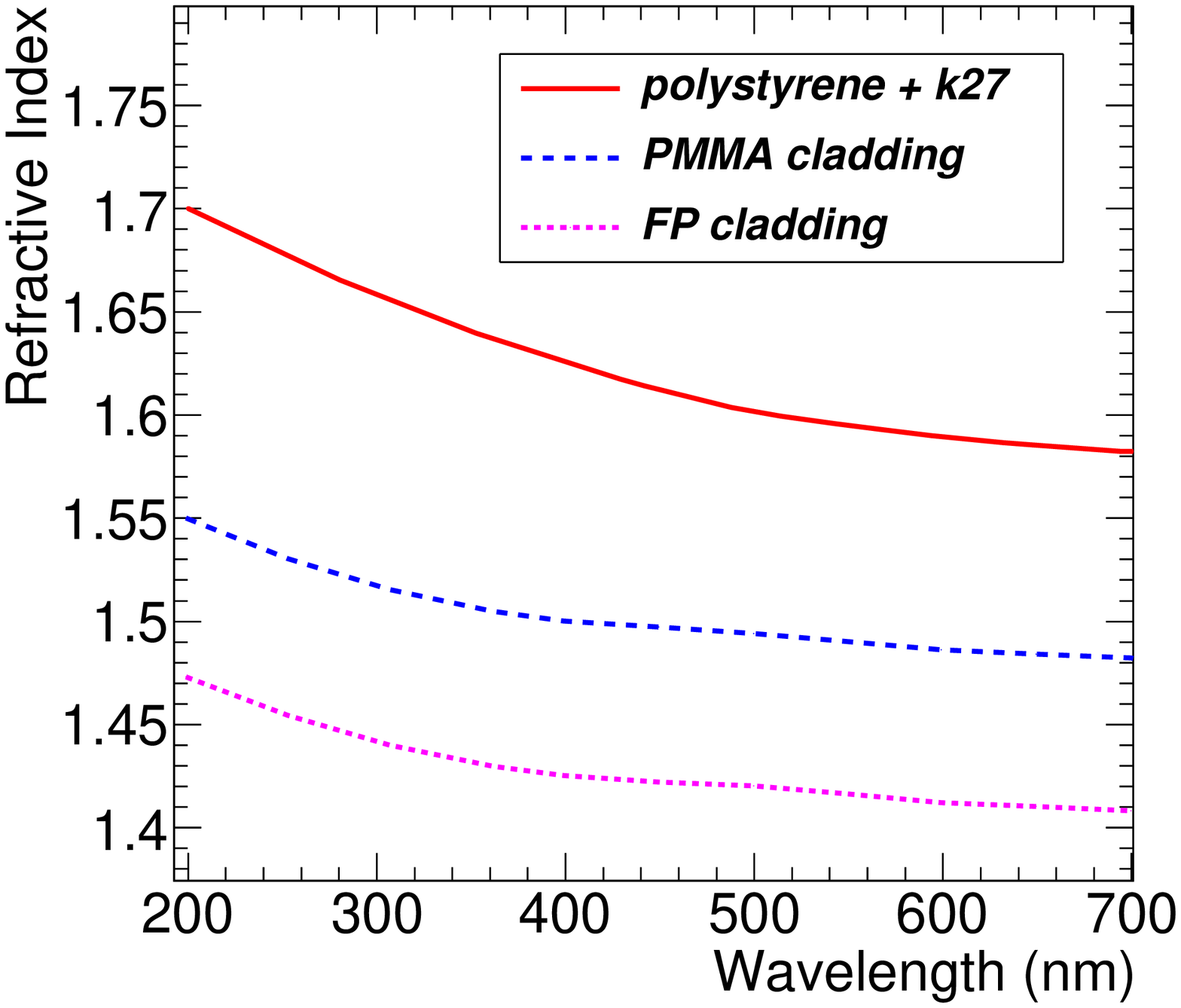}
}
\caption{Left: The emission spectrum of the dopant molecule K27 estimated from 
Kuraray~\cite{Kuraray_1} in the absence of bulk material absorption.  
Right: The refractive indices of the wavelength shifting fiber core and each cladding. }
\label{figure:k27_emission}
\end{figure}

Figure~\ref{figure:k27_emission} shows the emission spectrum of K27 in the absense of bulk absorption used as an input to the simulation.  The data have been estimated from~\cite{Kuraray_2}.  Figure~\ref{figure:k27_emission} also shows the refractive indices of the core and each cladding.  The data for the refractive index of the core and the claddings were taken from \cite{Kuraray_1} and \cite{Kasarova:2007}, respectively.

In all the simulated setups, we generate 10 million photons for propagation through the fibers.  For WLS fibers, a small fraction of these are absorbed, isotropically re-emitted, and propagated in the fiber after interaction with a WLS molecule.  We assume 100\% detection efficiency which is consistent with the corrected spectra from both the Ocean Optics spectrophotometer and the CCD camera.  The incident photons are distributed using a normal distribution across the width of the fiber with a transverse spread equal to the fiber radius including illumination of both claddings.


\section{Systematics}
 
There are several sources of systematic uncertainty in the data, largely attributed to the illumination methods and LEDs.  We evaluate the systematics using 1 meter fibers and considered several sources of fluctuation.  First, we varied the voltage applied to the LEDs from 2 to 3.5\,V.  This resulted in global shifts in the fiber spectra less than about 10\,nm, therefore we apply a 5\,nm shift as uncertainty.  We varied the pulse width from 1 to 20 microseconds which led to a global shift of about 5\,nm over this range, therefore we apply an uncertainty of 2.5\,nm.  We observed no effects due to changing the frequency.   

There are several sources of systematic uncertainty in the simulation model.  To account for variation in the bulk absorption spectra, we scaled the Y-11 absorption spectrum and the polystyrene bulk absorption spectrum individually by $\pm$10\%.  To account for uncertainty in the Y-11 emission spectrum, due to the available input data, we shifted the distribution by $\pm$5\,nm.  To account for variation and uncertainty in the fiber diameter we changed the nominal value of the fiber diameter by $\pm$0.1\,mm.  

\begin{table}
\centering
\begin{tabular}{ |l | c | c |}
\hline
Systematic & \% Fluctuation & Relative Difference from Nominal  \\
\hline
\hline
Y-11 absorption length        	& $\pm$ 10\%            	& 0.05       \\ \hline
Polystyrene absorption length 	& $\pm$ 10\%            	& 0.01       \\ \hline
K27 emission spectrum         	& $\pm$ 5\,nm            	& 0.06       \\ \hline
Fiber diameter                		& $\pm$ 0.1\,mm          	& 0.02       \\ \hline
\end{tabular}
\caption{Data and Monte Carlo systematic uncertainties. 
Relative difference from nominal represents total deviation in the total collected number of photons.  }
\label{table:systematics}
\end{table}

\section{Results}

\subsection{Validation with clear fibers}

\begin{figure}[h]
\begin{center}$
\begin{array}{cc}
\includegraphics[width=0.45\textwidth]{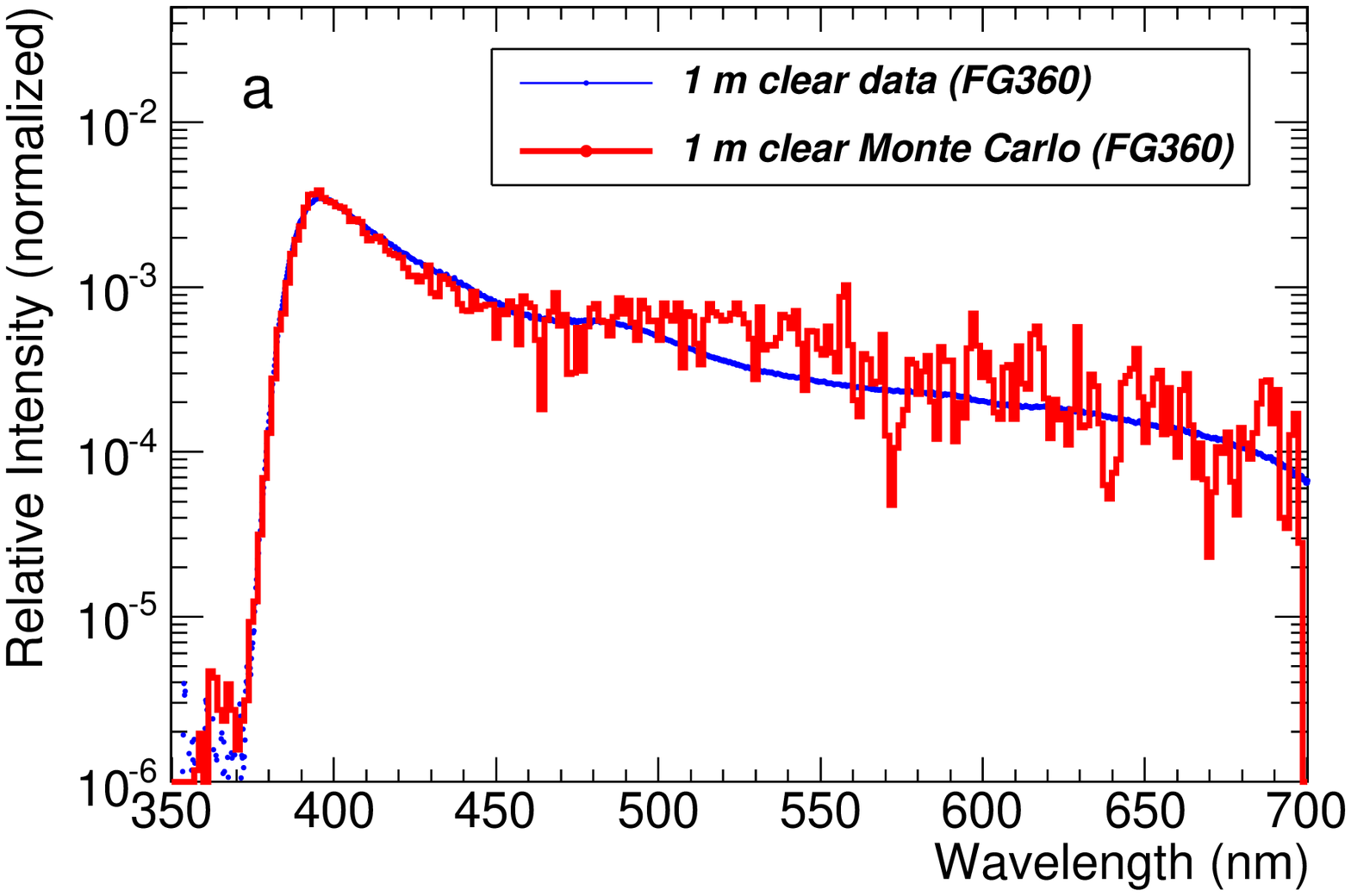} &
\includegraphics[width=0.45\textwidth]{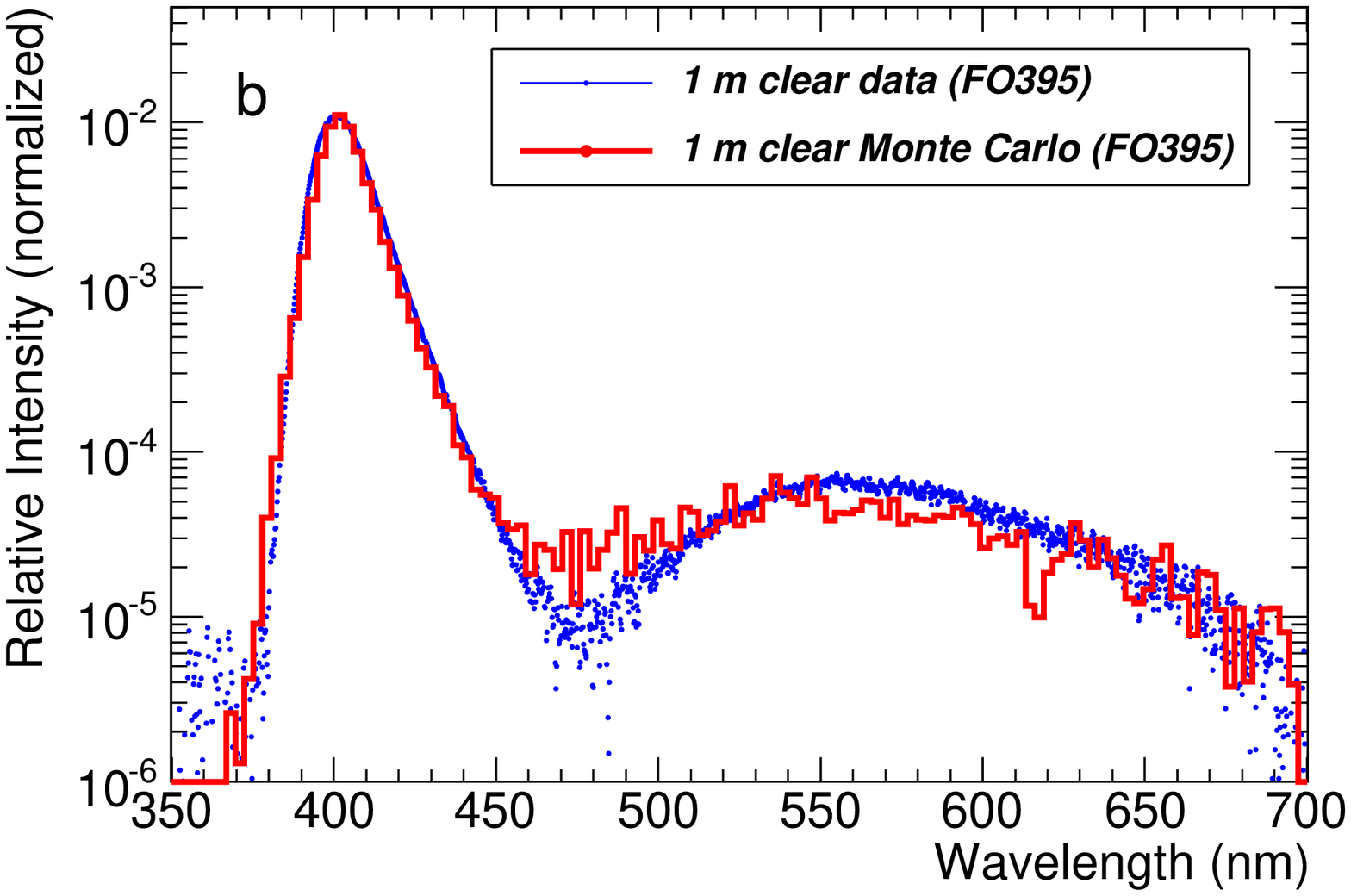} \\ 
\includegraphics[width=0.45\textwidth]{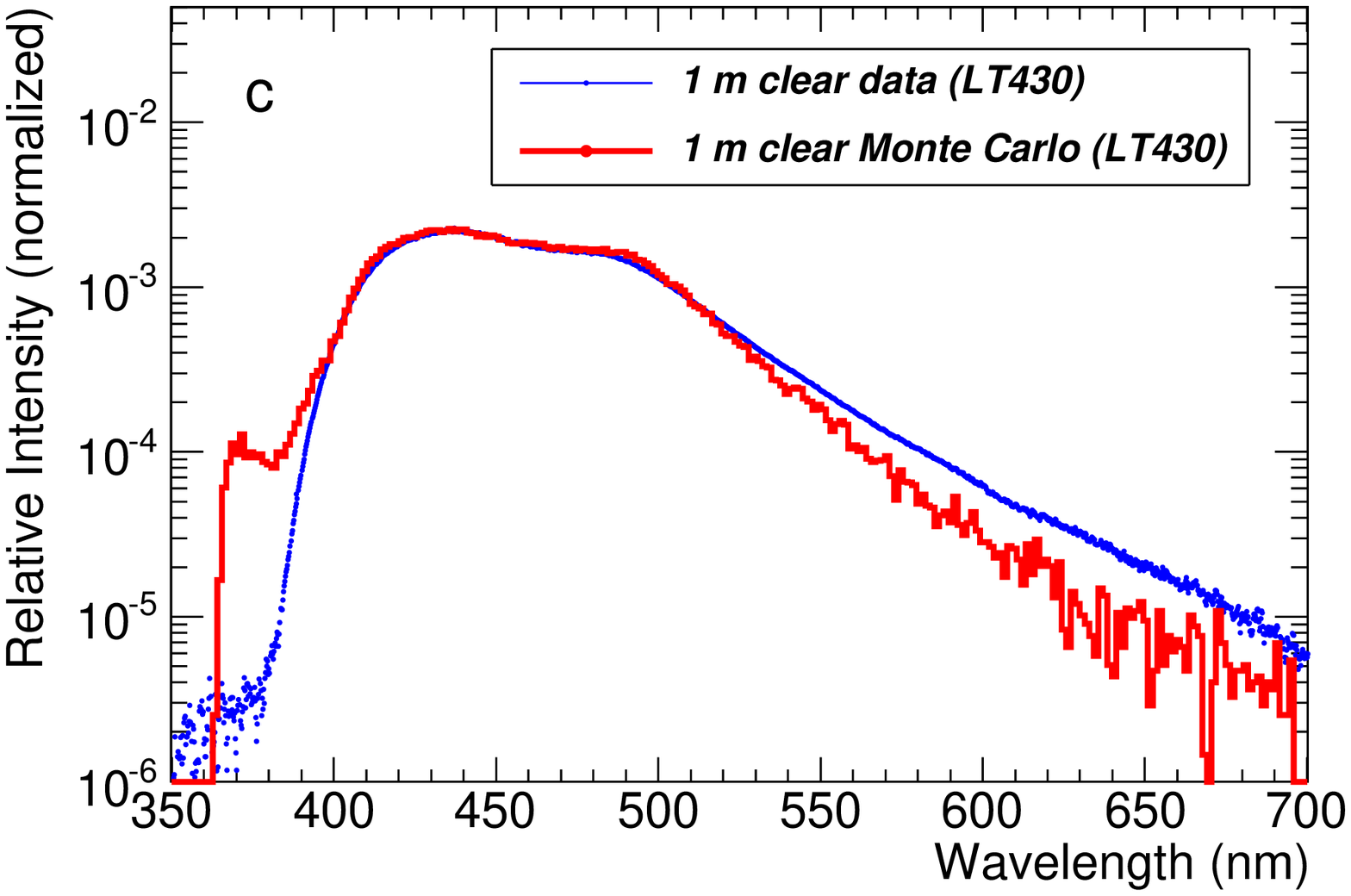} &
\includegraphics[width=0.45\textwidth]{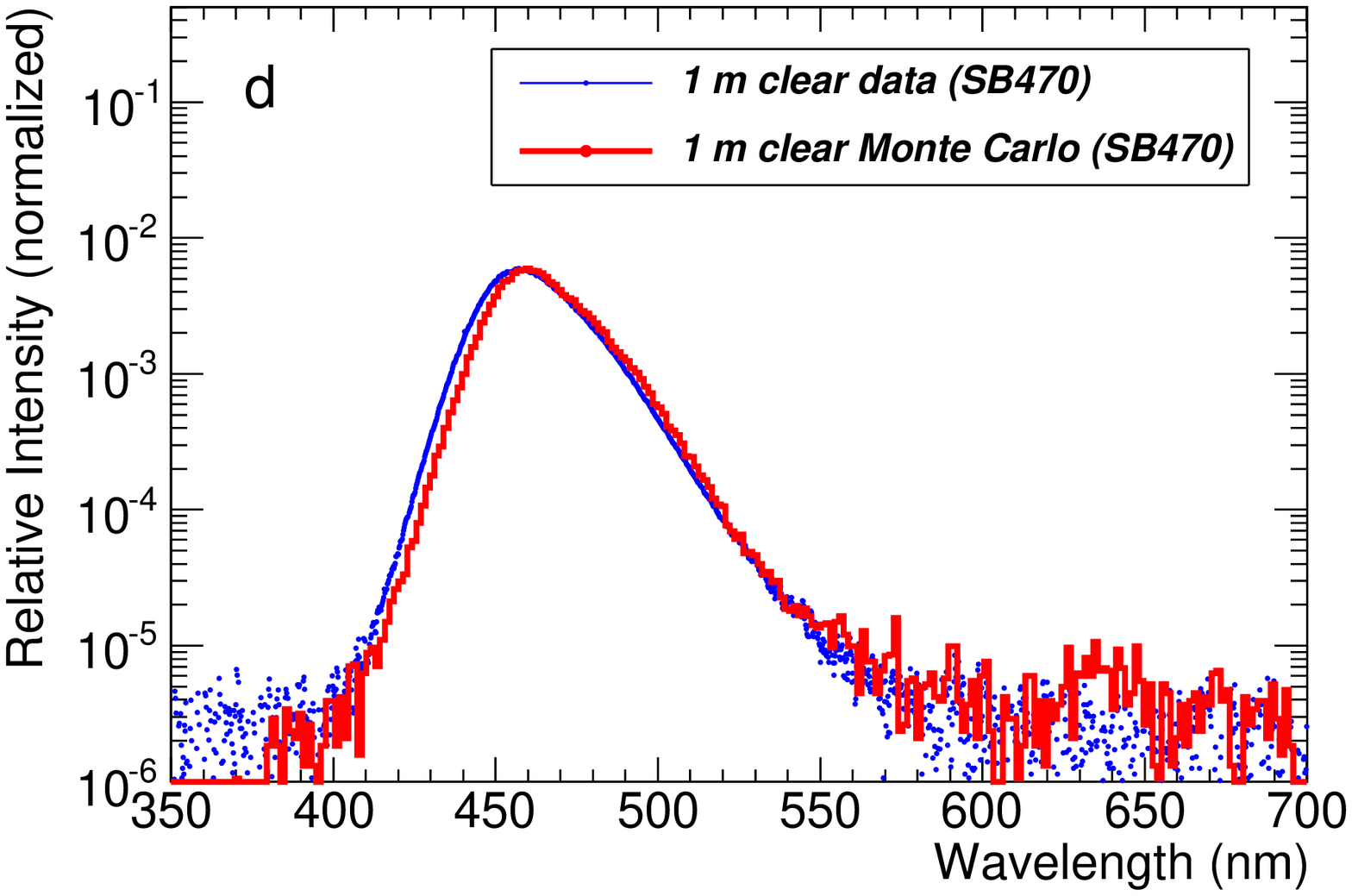} \\
\end{array}$
\end{center}
\caption{A comparison of data and simulation for a clear 1\,m 1.4\,mm diameter polystyrene fiber illuminated with four different LEDs.  
The thin line shows the data obtained from the spectrophotometer and the thick line shows the simulation.  The distributions are shown for illumination with a) 360\,nm LED, b) 395\,nm LED, c) 430\,nm LED and d) 470\,nm LED. }
\label{figure:all_leds_on_clear_fibers}
\end{figure}

We considered decoupling the effects of the dopant molecule by first studying the propagation and attenuation of light in clear fibers.  Figure~\ref{figure:all_leds_on_clear_fibers} shows the data and Monte Carlo spectra obtained for illumination of a clear 1\,m  long 1.4\,mm diameter polystyrene fiber using four LEDs.  We illuminated the face of the fiber with collimated light to allow for direct propagation.  The four LEDs probe most of the region of interest between 350\,nm and 600\,nm.  The results show good agreement for all four LEDs and indicate sufficient knowledge of the clear fiber spectral attenuation length.  This will be used in the subsequent simulations.  

\subsection{Validation with WLS fibers}

\begin{figure}[h]
\centerline{
\includegraphics[width=0.9\linewidth]{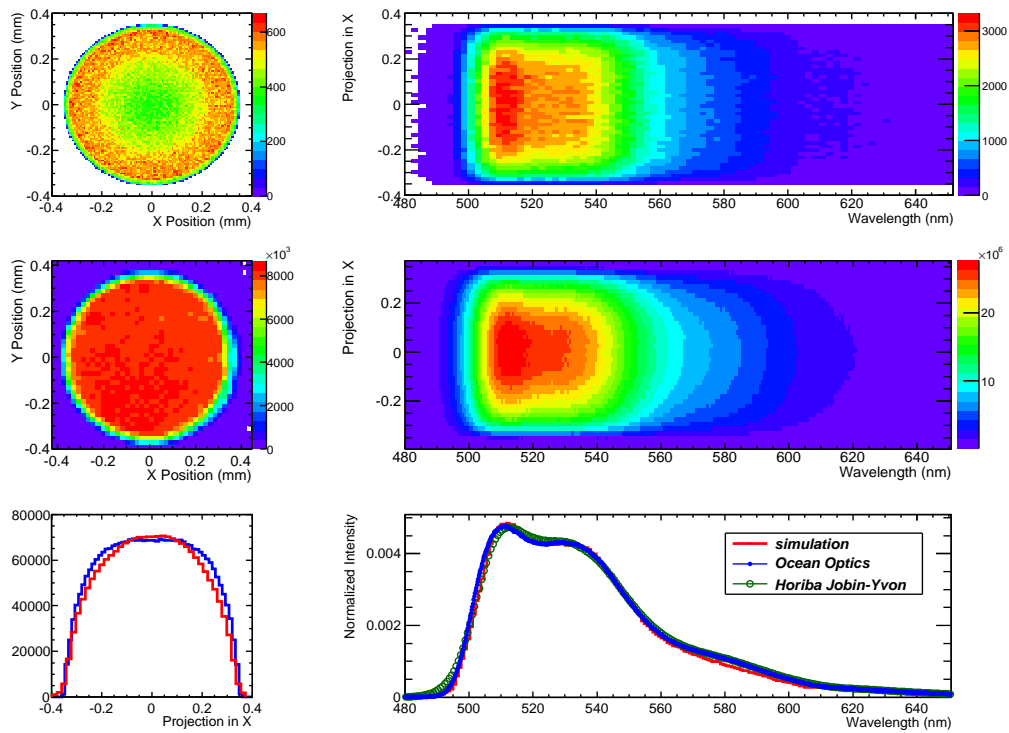}
}
\caption{Comparison of the exit position and wavelength spectra for data and Monte Carlo for a 200\,ppm, 0.7\,mm diameter, 1\,m fiber.  The top left shows the simulated exit position with respect to the fiber cross section.  The top right shows the simulated exit position as a function of wavelength. The center left shows the measured exit position with respect to the fiber cross section.  The center right shows the measured exit position as a function of wavelength, both obtained from the Horiba spectrometer/CCD. The bottom left shows a comparison of data and simulation for the projection of the exit position onto the X-axis.  The bottom right shows a comparison of data (for both the Horiba spectrometer/CCD and the Ocean Optics spectrophotometer) and simulation for the wavelength distribution.  }
\label{figure:200ppm_spectrum_0.7}
\end{figure}

We measured the spectral and spatial output at the end of a 1 meter long, 0.7\,mm diameter, 200\,ppm WLS fiber using two different acquisition systems and compared the result to simulation.  The results of this comparison are shown in Figure~\ref{figure:200ppm_spectrum_0.7}.  The top two figures show the results of the simulation and the middle set shows the measured data for the spatial and spectral distributions of exiting photons.  The bottom set of figures shows the projection of the spatial distribution onto the X-axis and the spectral distribution onto the wavelength axis.  For the spectral distribution, we show a comparison of the Monte Carlo results to the data obtained from both the Ocean Optics spectrophotometer and the Horiba Jobin-Yvon spectrometer.  The spatial projections in data are uniformly flat while those for Monte Carlo show a trend towards more photons arriving at the edges.  We investigated this difference in several ways.  First, we considered the possibility that the difference arose from the illumination scheme.  We simulted photons in a pencil beam directed at the center of the fiber and we simulated photons in a uniform distribution across the diameter of the fiber.  Both cases resulted in a similar output distribution.  We also considered the possibility that the fiber possibly had a graded refractive index.  Here, we delineated the core into ten cylindrical shells, each with an individual refractive index going radially outward from 1.55 to 1.65.  This also resulted in a similar distribution.  Finally, we increased the Rayleigh scattering attenuation length by a factor of two, giving similar results as well.  We speculate that this difference arises from the notion that the light propagation modes in this multi-mode fiber are ideal and that the mode mixing is minimal.  The wavelength distribution, on the other hand, shows very good agreement between the two sets of data and both agree well with Monte Carlo.

\subsection{Measurements of the spectral evolution through WLS fiber} 

\begin{figure}[h]
\begin{center}$
\begin{array}{cc}
\includegraphics[width=0.40\textwidth]{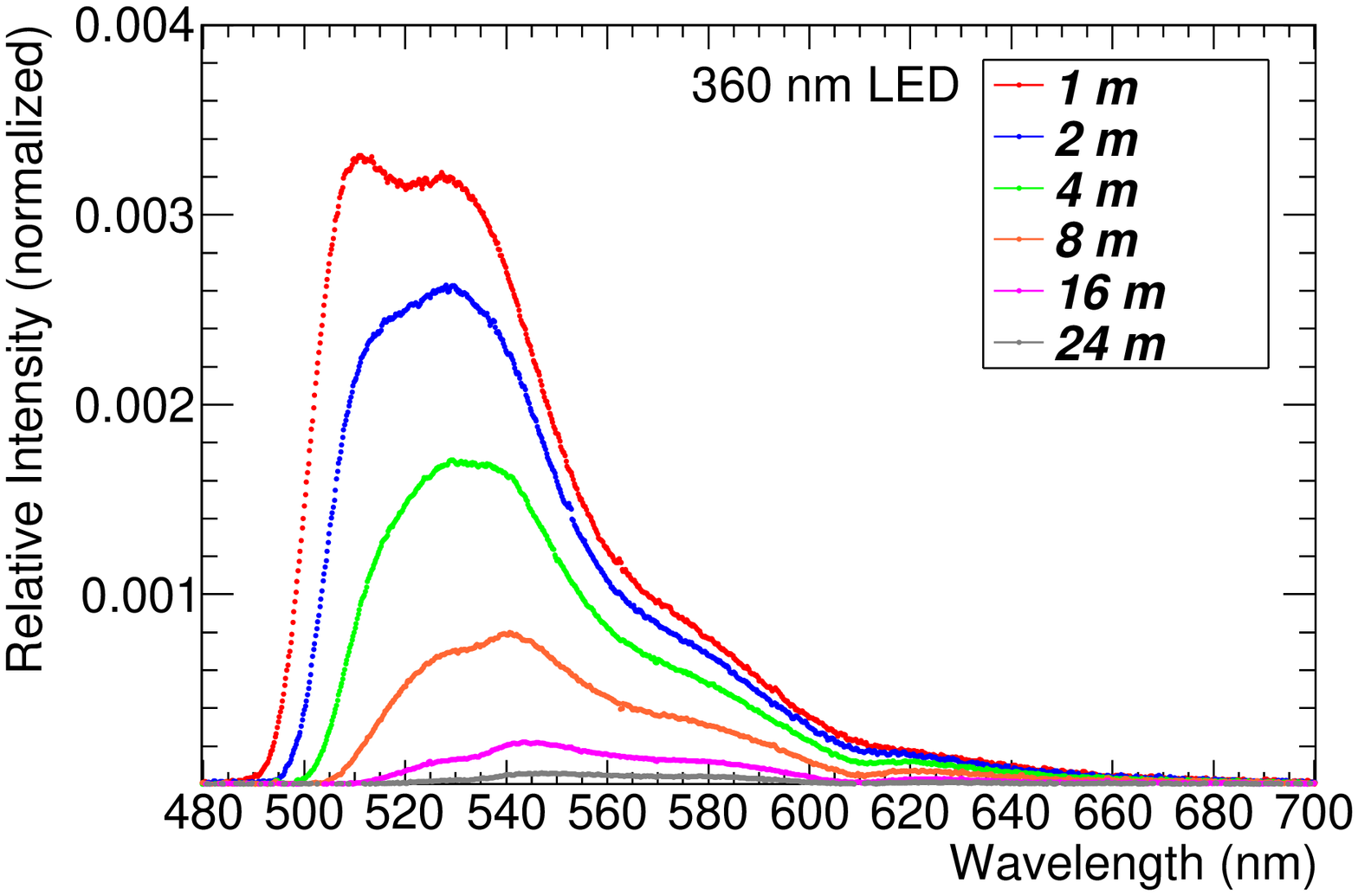} &
\includegraphics[width=0.40\textwidth]{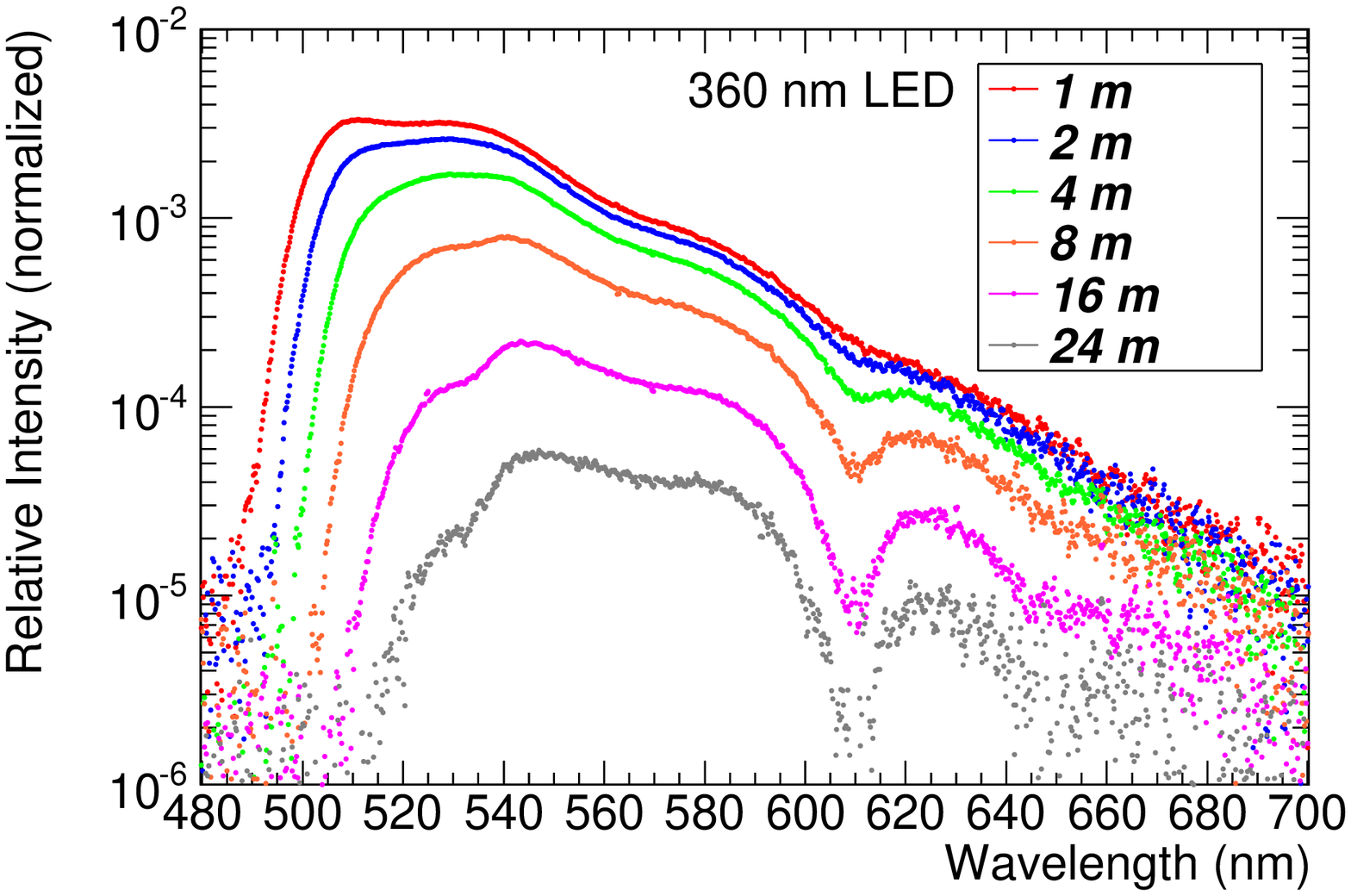} \\ 
\includegraphics[width=0.40\textwidth]{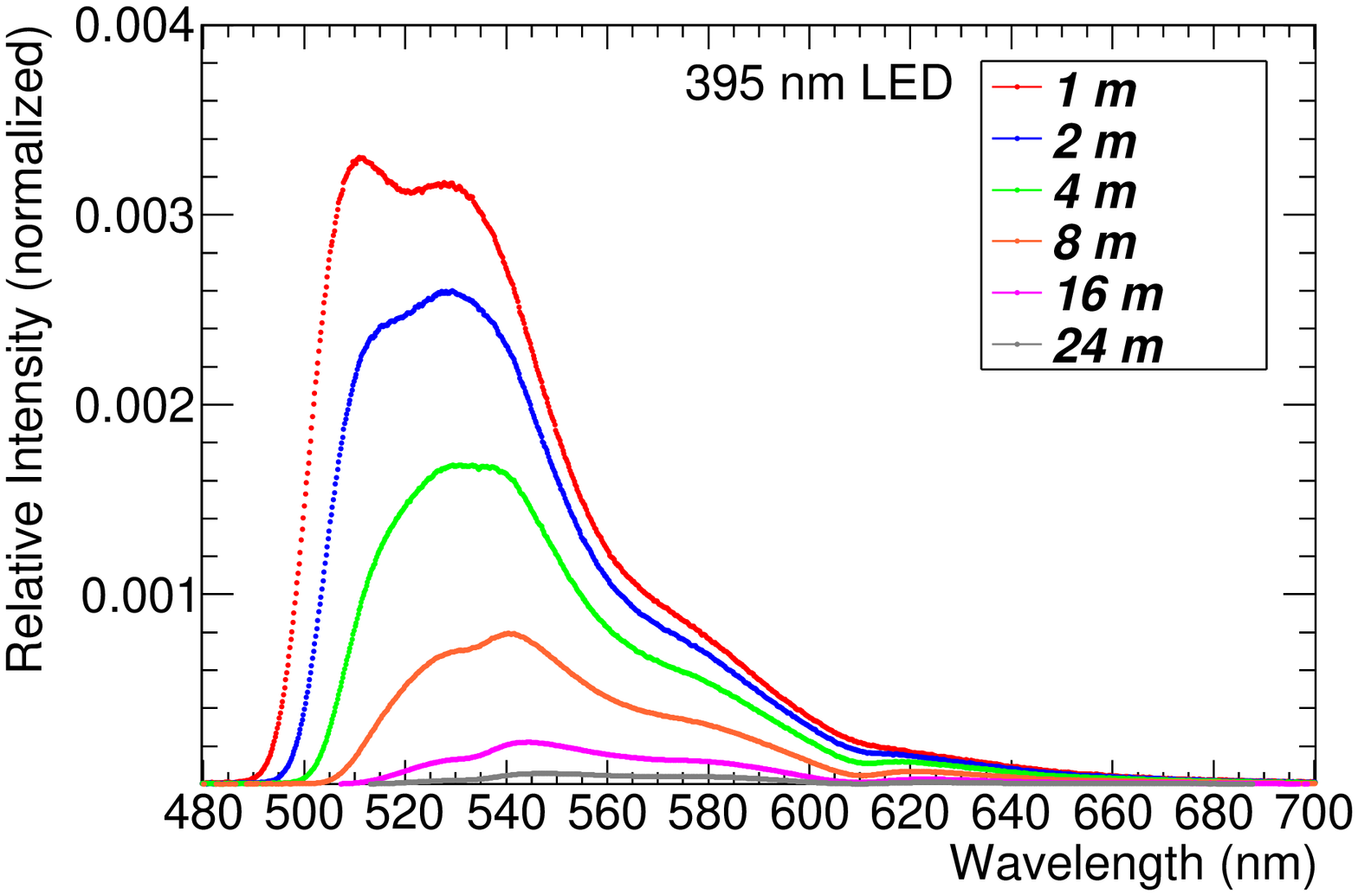} &
\includegraphics[width=0.40\textwidth]{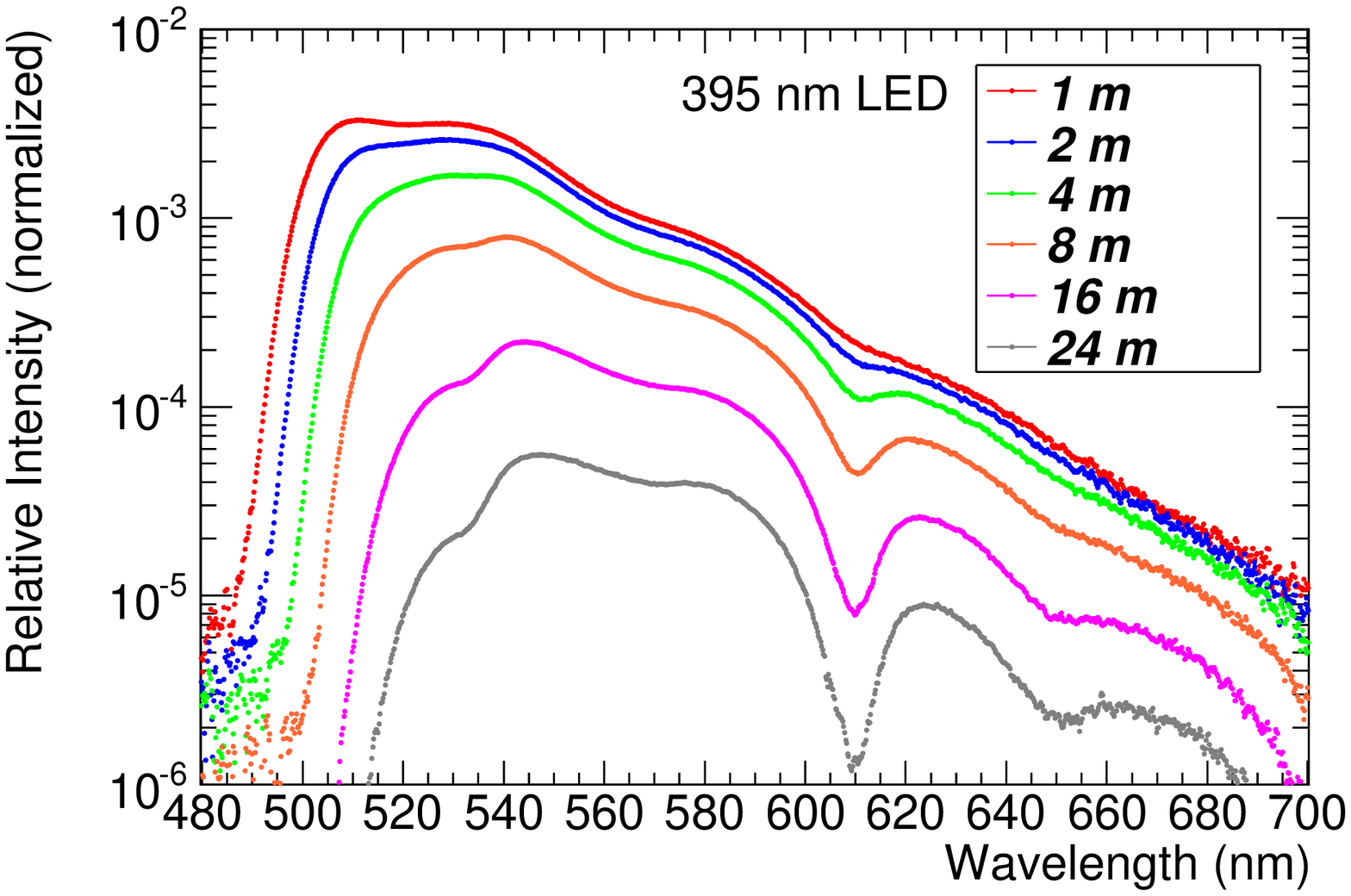} \\
\includegraphics[width=0.40\textwidth]{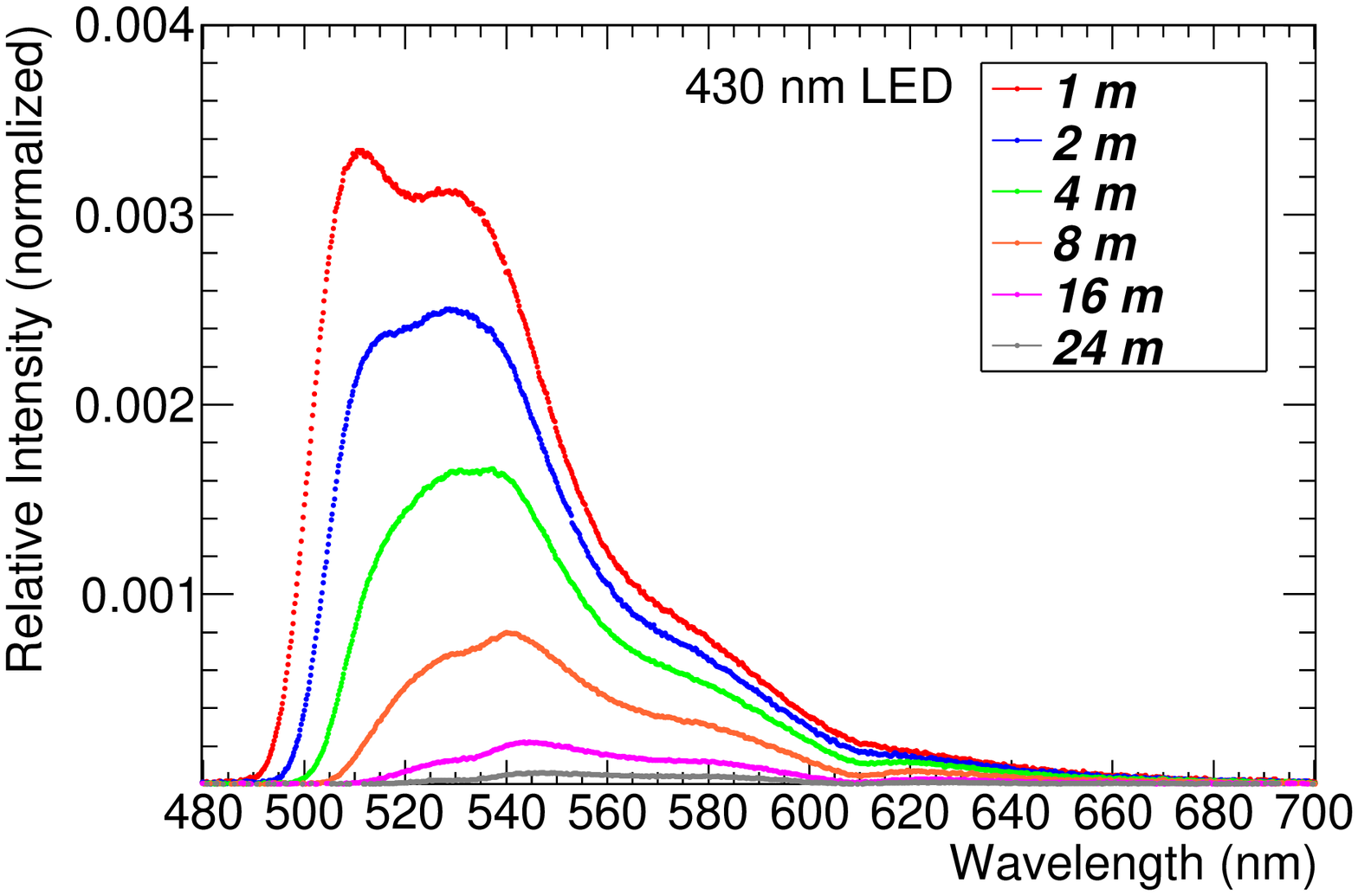} &
\includegraphics[width=0.40\textwidth]{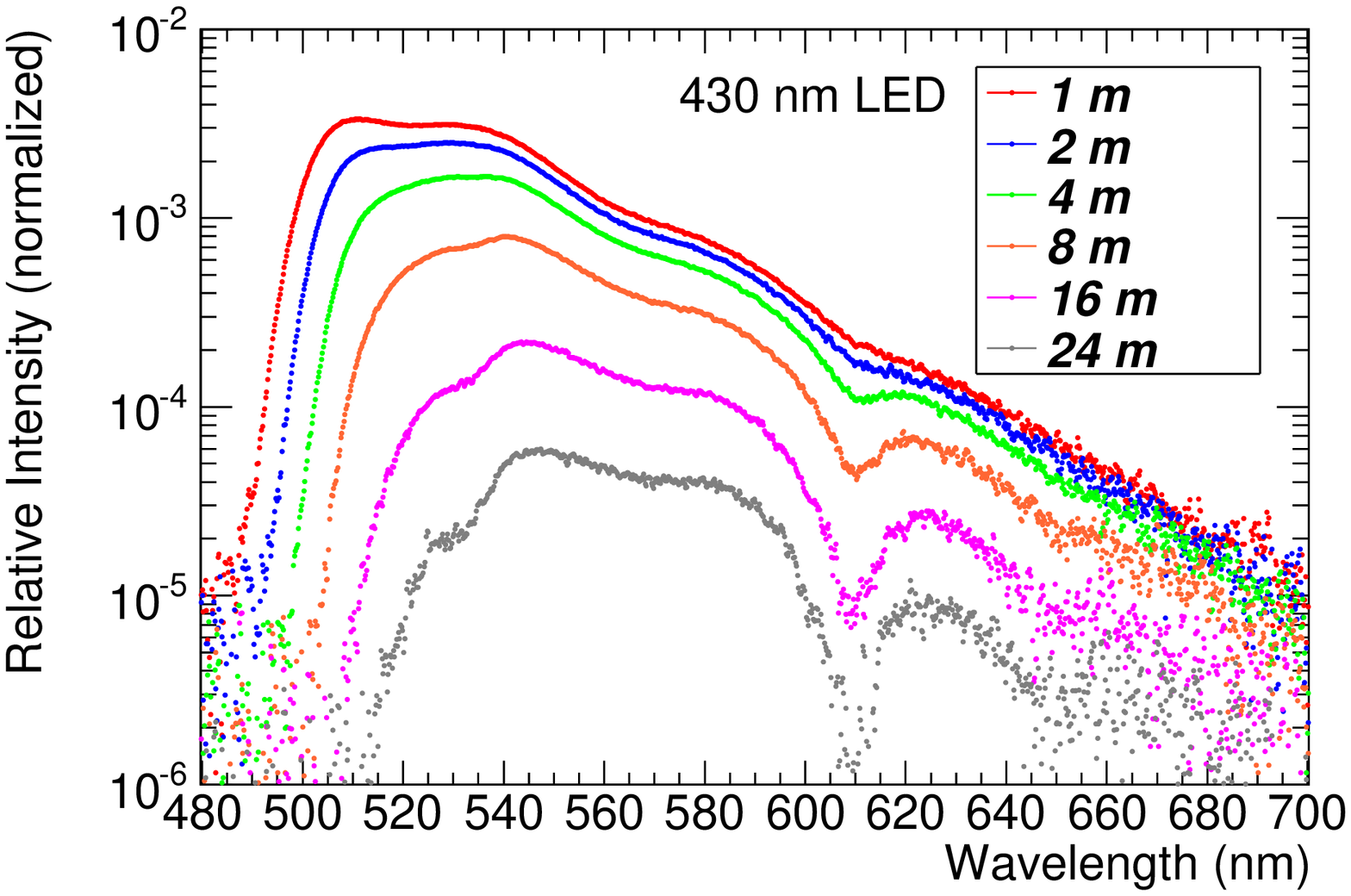} \\ 
\includegraphics[width=0.40\textwidth]{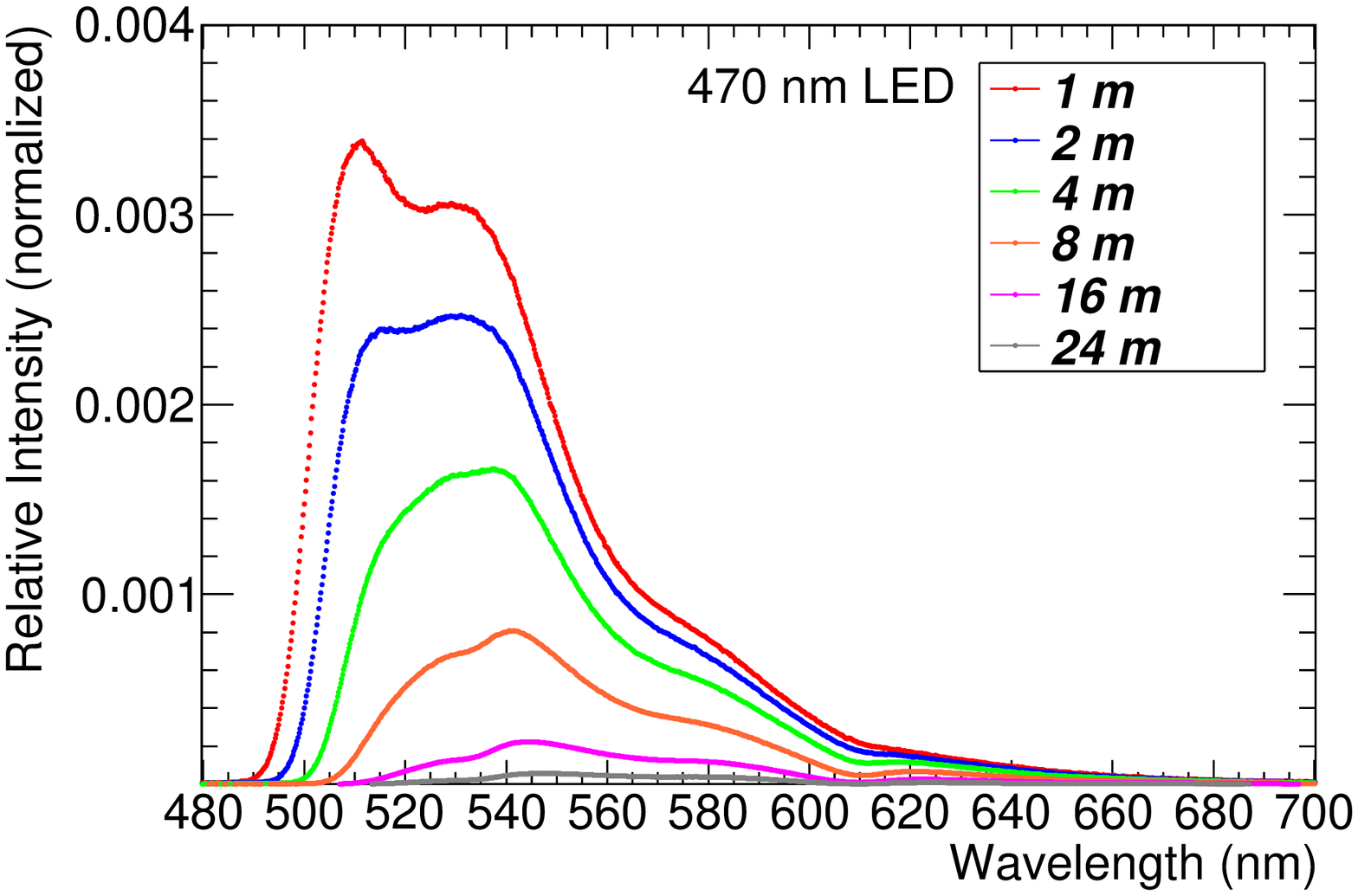} &
\includegraphics[width=0.40\textwidth]{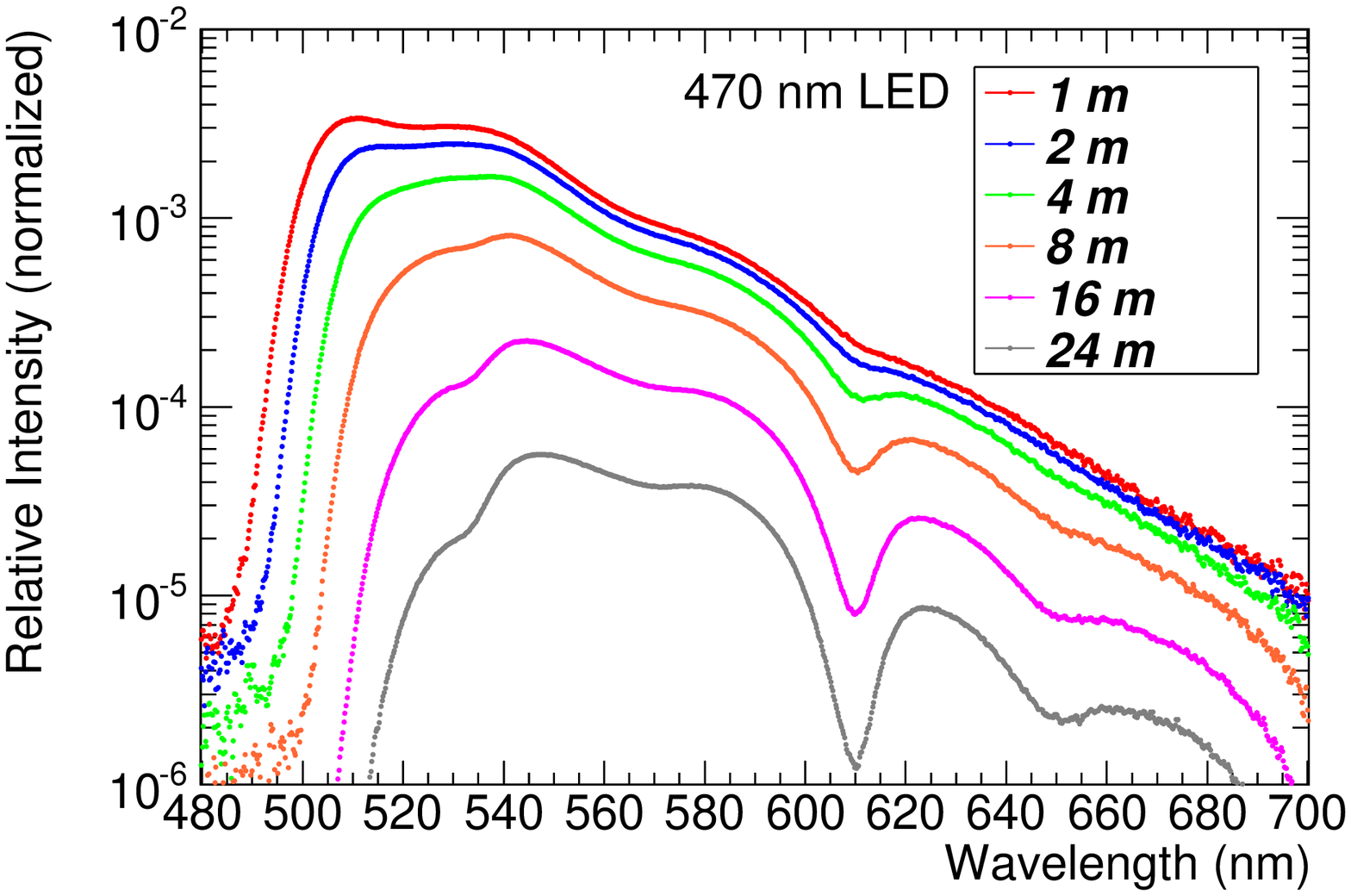} \\
\end{array}$
\end{center}
\caption{Spectra obtained from 1 to 24 meters for 0.7\,mm-diameter WLS fiber illuminated perpendicular to the principal fiber axis using four different LEDs (See Figure~\ref{fig:led_spectra}).  The linear distributions are shown on the left and the log distributions are shown on the right.  Each distribution is normalized to unit area then normalized to the acquisition time.}  
\label{figure:all_leds_on_fibers}
\end{figure}

To investigate the evolution of spectra during propagation through the fiber, various lengths 1--24 meters of 0.7~mm-diameter Kuraray Y11(200) WLS fibers were illuminated with LEDs directed on the side of the fiber. Figure~\ref{figure:all_leds_on_fibers} show the spectra of light after propagation through several lengths of fiber.     All curves are normalized to collection time then normalized to unit area.  The distributions with respect to each length are very similar, which suggests that the spectral distributions at the end of each fiber is independent of illumination at different wavelengths.  The dip at 525\,nm arises from two independent effects.  At short fiber lengths (e.g 1 meter), the dip is attributed to the K27 emission spectrum.  However, as shorter wavelengths are suppressed, the dip arises due to an absorption band at this wavelength region in the polystyrene.  The dips at 570, 610, and 650\,nm at long fiber lengths are also attributed to absorption bands in the polystyrene.  We do not observe any features consistent with absorption bands in the PMMA; this suggests that the cladding plays a minimal role in the overall absorption and resulting spectra.  Suggested previously, the properties of the core material dictate the overall attenuation since the optical field does not penetrate very far into the cladding~\cite{Emslie}.

\begin{figure}[h]
\centerline{
\includegraphics[width=0.45\linewidth]{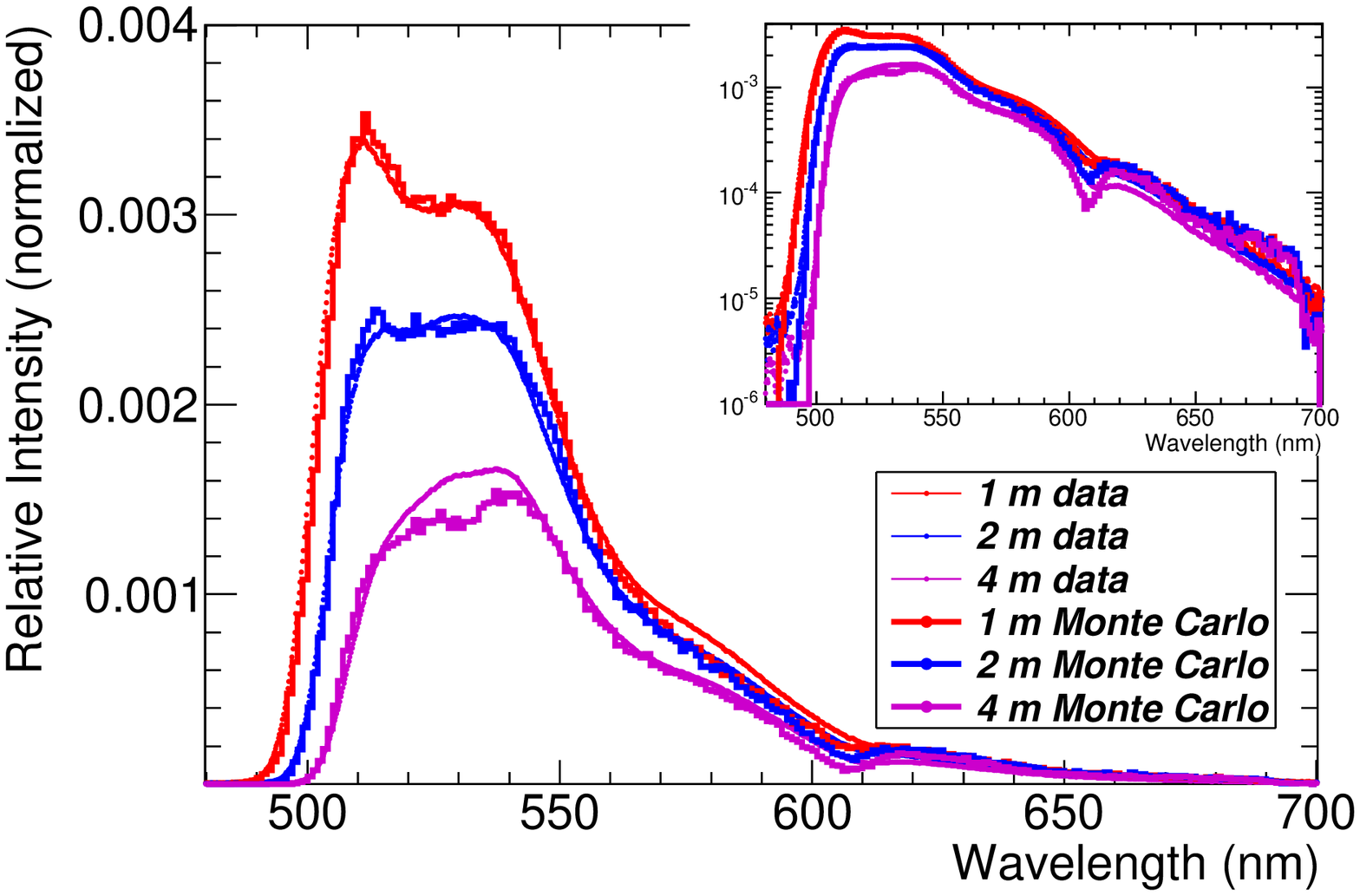}
\hskip0.25in
\includegraphics[width=0.45\linewidth]{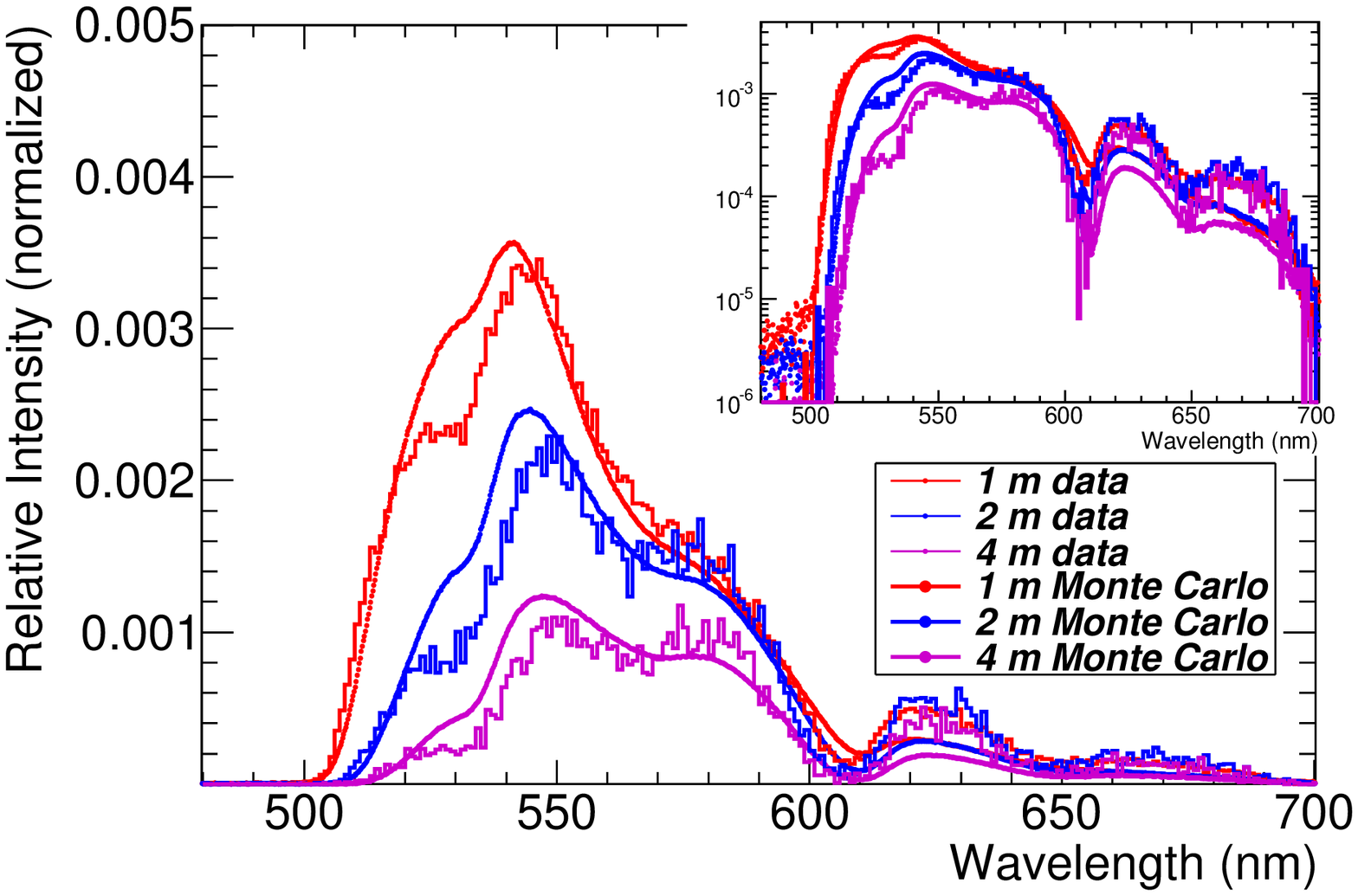}
}
\caption{Wavelength spectra comparison of data and Monte Carlo for a 200\,ppm 0.7\,mm diameter WLS fiber.  The data is shown as dots and the Monte Carlo is represented by histograms.  Left: Results for 1, 2, and 4 meter fiber lengths. Right: Results for 8, 16, and 24 meter fiber lengths.   }
\label{figure:1_to_24_meter_dataMC}
\end{figure}

We further validated the model by studying the evolution of the wavelength distribution for several lengths of WLS fiber by holding the input data constant.  Figure~\ref{figure:1_to_24_meter_dataMC} shows a comparison of the wavelength spectra for data and Monte Carlo for a 0.7\,mm diameter, 200\,ppm WLS fiber for several lengths.   There is good agreement for shorter fibers, especially in the 500--530\,nm range.  Small disagreements are seen for longer fiber lengths.  The deficit in Monte Carlo near the 525\,nm region is due to an underestimation of the absorption length.  Towards longer wavelengths, the enhancement is due to uncertainty in the K27 emission spectrum.  This demonstrates the ability of the model to reproduce the data with no additional tuning.

\subsection{Spectral evolution through WLS fiber:  data--MC comparison}

\begin{figure}[h]
\centerline{
\includegraphics[width=0.45\linewidth]{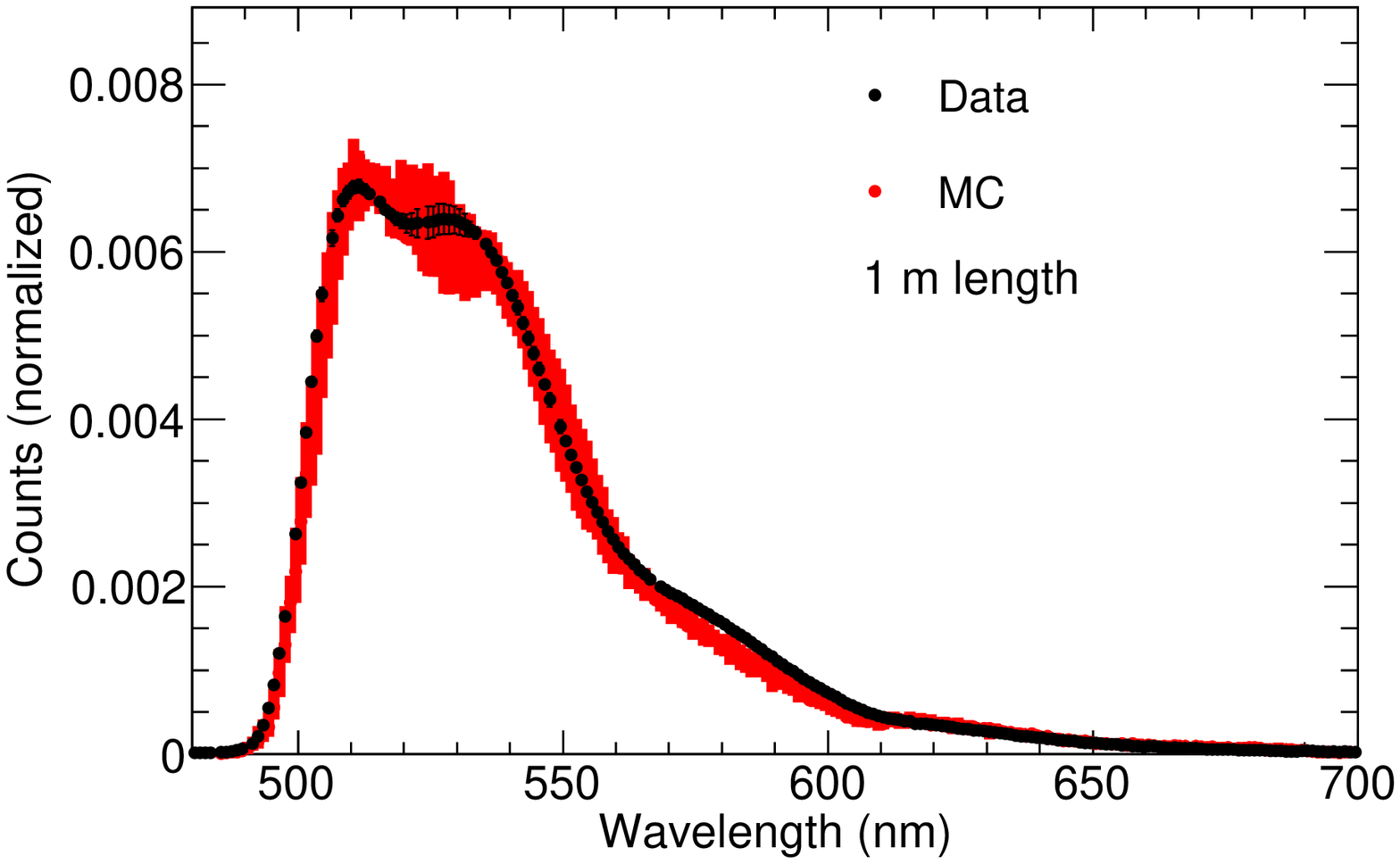}
\hskip0.25in
\includegraphics[width=0.45\linewidth]{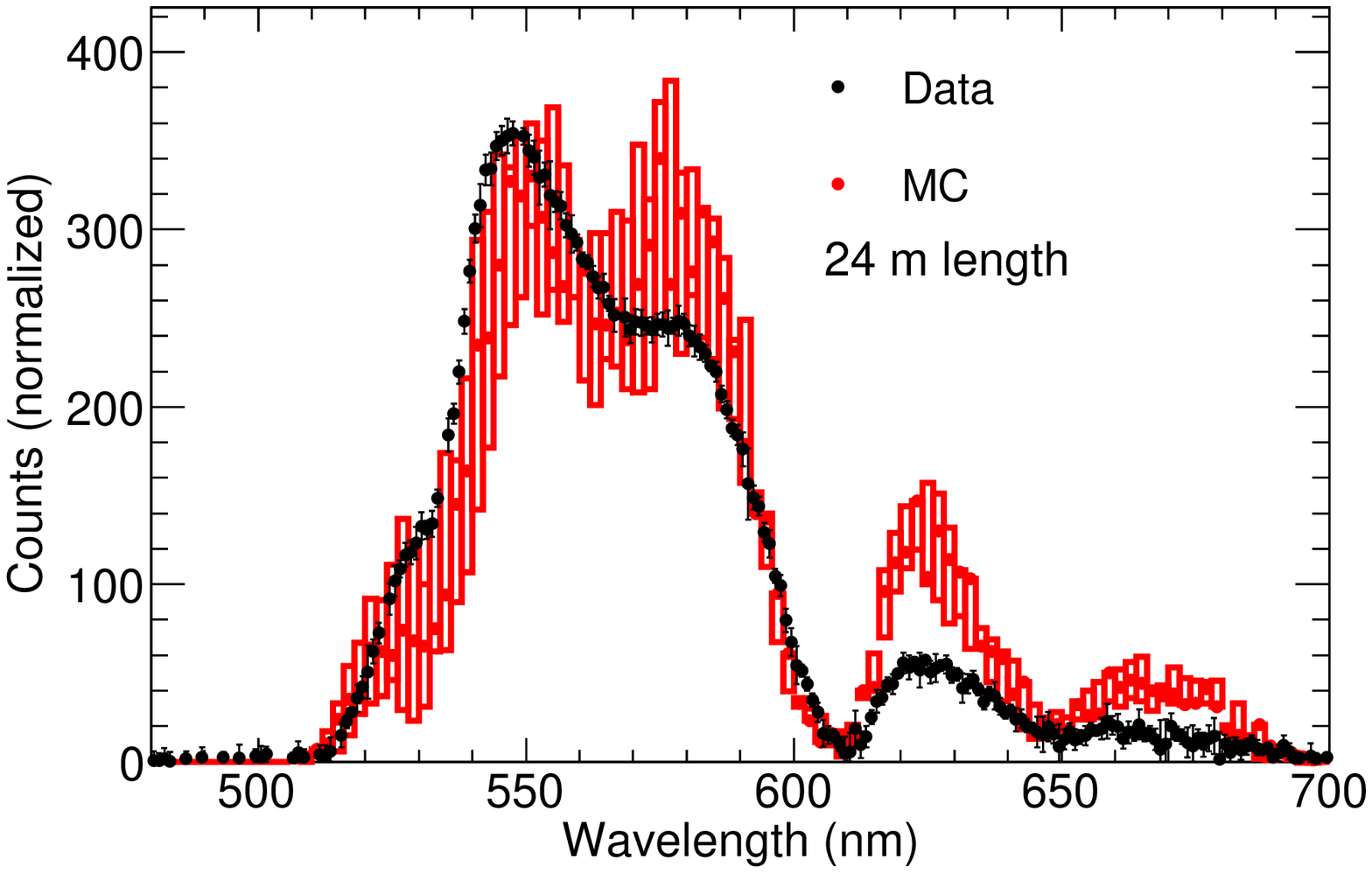}
}
\caption{Wavelength spectra uncertainties for a 200\,ppm 0.7\,mm diameter WLS fiber.  Left: Results for a 1 meter length fiber.  Right: Results for a 24 meter length fiber.   The black points represent measured data with uncertainty due to LED stability and LED wavelength variation, the red points represent the nominal Monte Carlo and the red histogram represents Monte Carlo estimated systematic error bands.   }
\label{figure:1m_data_mc_wsyst}
\end{figure}

Figure~\ref{figure:1m_data_mc_wsyst} shows a comparison of data and Monte Carlo for a 0.7\,mm diameter, 200\,ppm, 1 meter length and 24 meter length fibers. The systematic error bands are generated by varying the input parameters as discussed previously.  For the 1m fiber, there is good agreement over all wavelengths, particularly the 500-550\,nm region where competing effects due to the two absorption lengths are present.  The sharp rise is due to the large attenuation below 500\,nm.  For the 24\,m fiber, the absorption bands are clearly evident at 530, 570, 610, and 650\,nm.  The systematic errors indicate a wider variation as more features become present at longer lengths.  The disagreements are due to both overcompensation of absorption band modeling combined with uncertainty in the K27 emission spectrum.

\subsection{Attenuation of overall light level in WLS fiber}

\begin{figure}[h]
\centering
\includegraphics[width=0.45\textwidth]{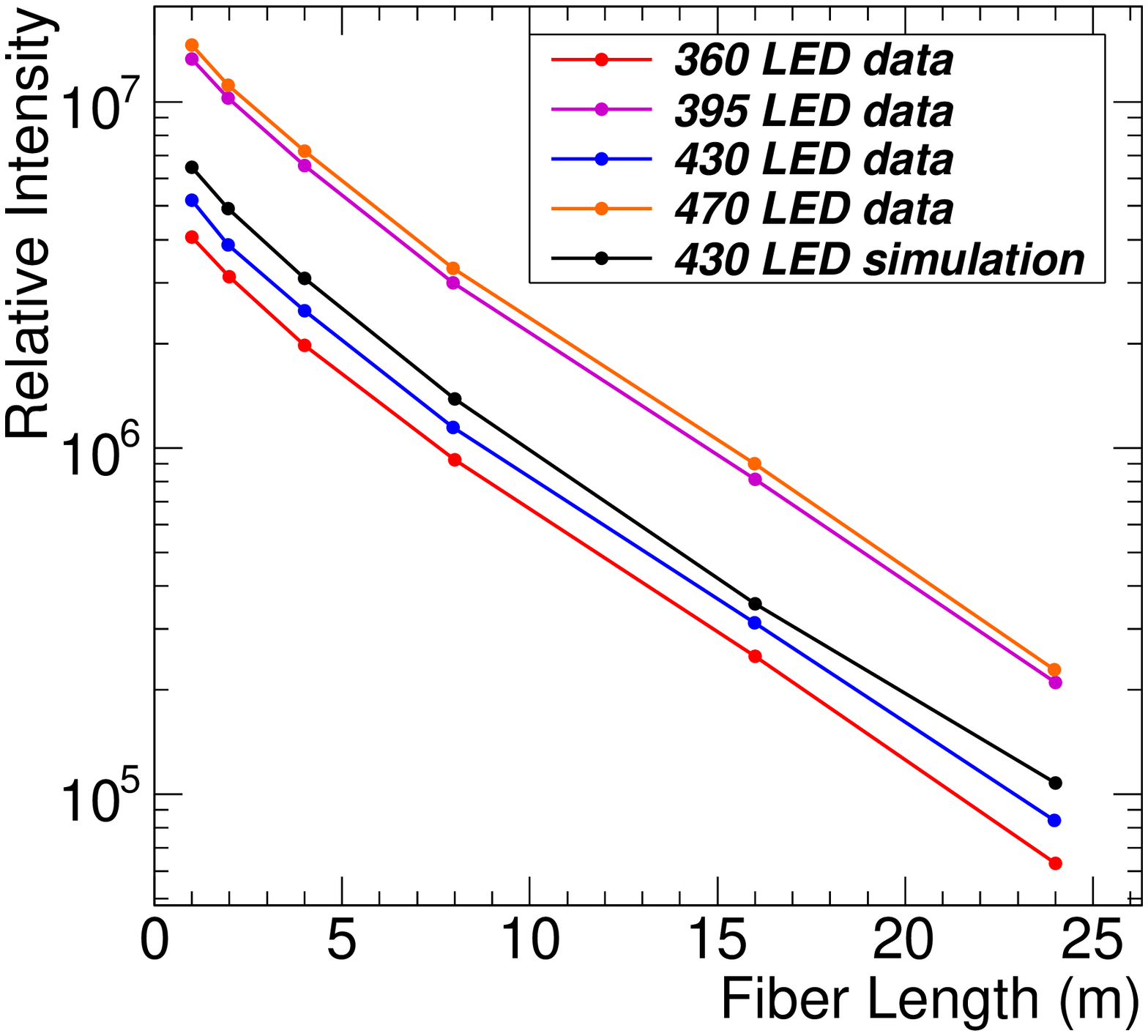} 
\hskip0.25in
\includegraphics[width=0.45\textwidth]{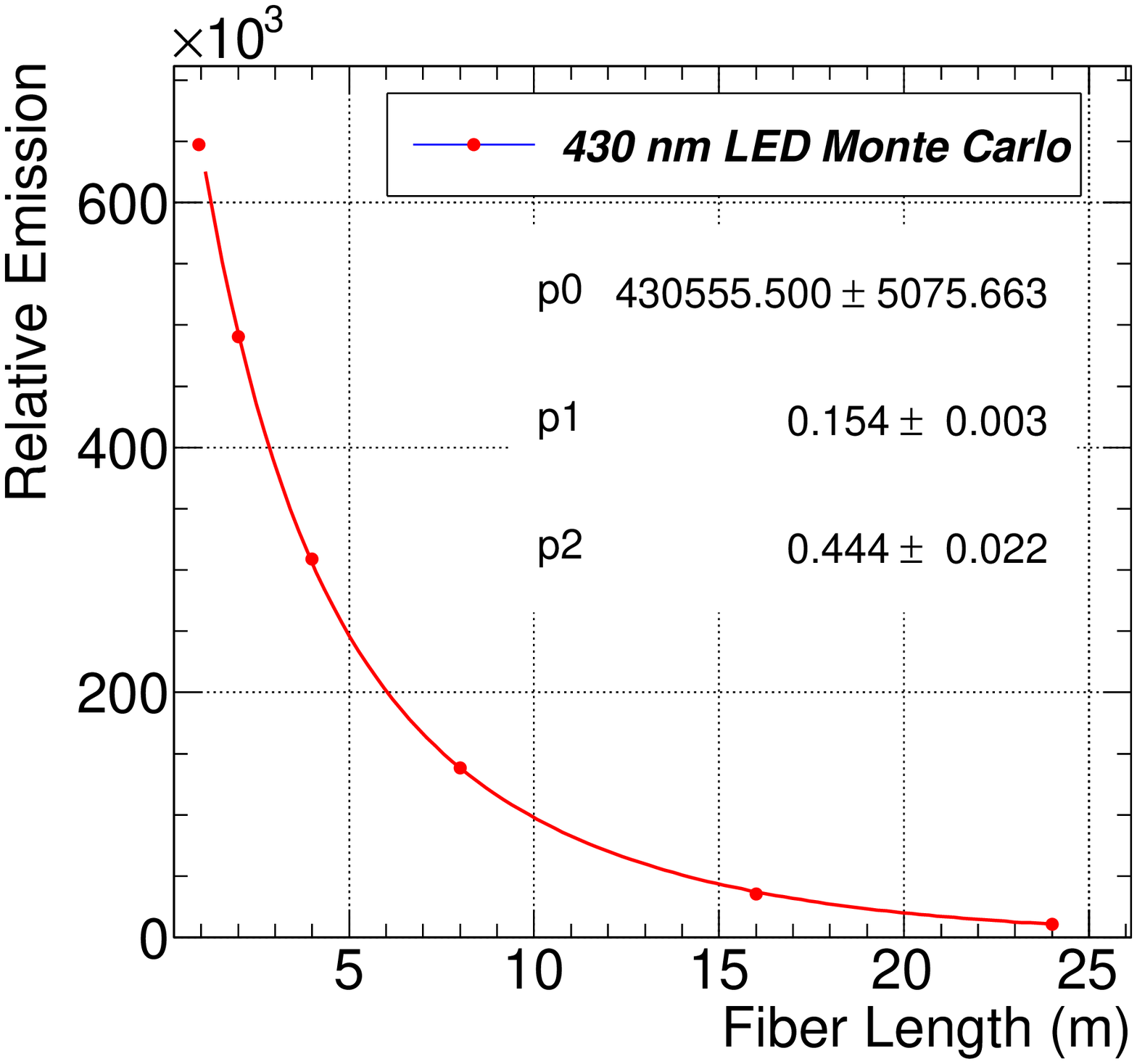} 
\caption{Left:  Attenuation of light through several lengths of 0.7\,mm-diameter, 200\,ppm WLS fiber, obtained by wavelength-integration of collection time-normalized spectra for data, and wavelength-integration for the 430\,nm LED Monte Carlo simulation. Right: A fit to a double exponential for simulation of the 430\,nm LED. }
\label{figure:atten}
\end{figure}

The spectra obtained for the 1 to 24\,m measurements and the simulation at 470\,nm were integrated to find the total transmitted intensity at each fiber length. Shown in Figure~\ref{figure:atten}, the attenuation curves roughly follow a pattern of exponential decay. However, it appears that light is attenuated more rapidly at short fiber lengths than at longer fiber lengths. This may be explained by the greater interaction of the fluorophores with shorter wavelength light, which is gradually shifted toward longer wavelengths as light propagates farther through the fiber. The attenuation was fit to a curve of the form $a(e^{-\lambda_1x} + e^{-\lambda_2x})$ as suggested in~\cite{Avvakumov:2005ww}.  The resulting attenuation lengths $\lambda_1$ and $\lambda_2$ for each curve are given in Table~\ref{tbl:atten-coeff-side}. The error estimates are produced by the fit.  There is good agreement between data sets.  Comparison of data and Monte Carlo using the 430\,nm LED also shows good agreement.

\begin{table}
\centering
\begin{tabular}{ |l | c | c | c | c | c | c |}
\hline
Diode & $\lambda_1$ (m) & $\lambda_2$ (m)  \\
\hline
\hline
FG360    	& 6.80 $\pm$ 0.09     & 2.38 $\pm$ 0.07        \\ \hline
FO395    	& 6.76 $\pm$ 0.13     & 2.43 $\pm$ 0.10        \\ \hline
LT430    	& 6.85 $\pm$ 0.23     & 2.13 $\pm$ 0.18        \\ \hline
SB470    	& 6.85 $\pm$ 0.18     & 2.32 $\pm$ 0.16        \\ \hline
LT430 MC & 6.49 $\pm$ 0.12     & 2.25 $\pm$ 0.08        \\ \hline

\end{tabular}
\caption{Attenuation rates for several different lengths of fiber illuminated with four LEDs 
and for the model prediction using a 430\,nm peak input spectrum. }
\label{tbl:atten-coeff-side}
\end{table}


\subsection{Dependence on LED illumination and orientation}

\begin{figure}[h]
\centering
\includegraphics[width=0.45\textwidth]{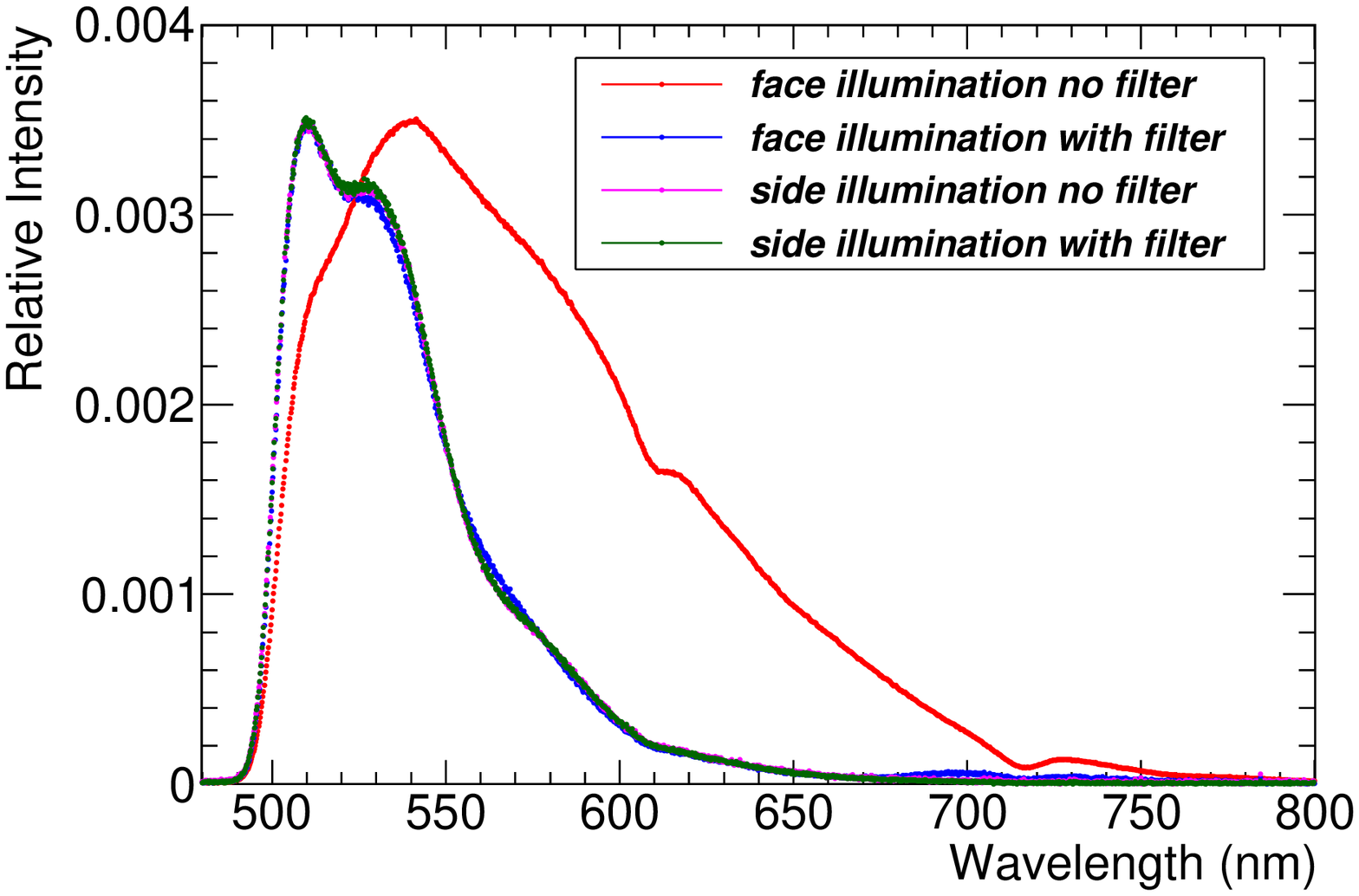} 
\hskip0.25in
\includegraphics[width=0.45\textwidth]{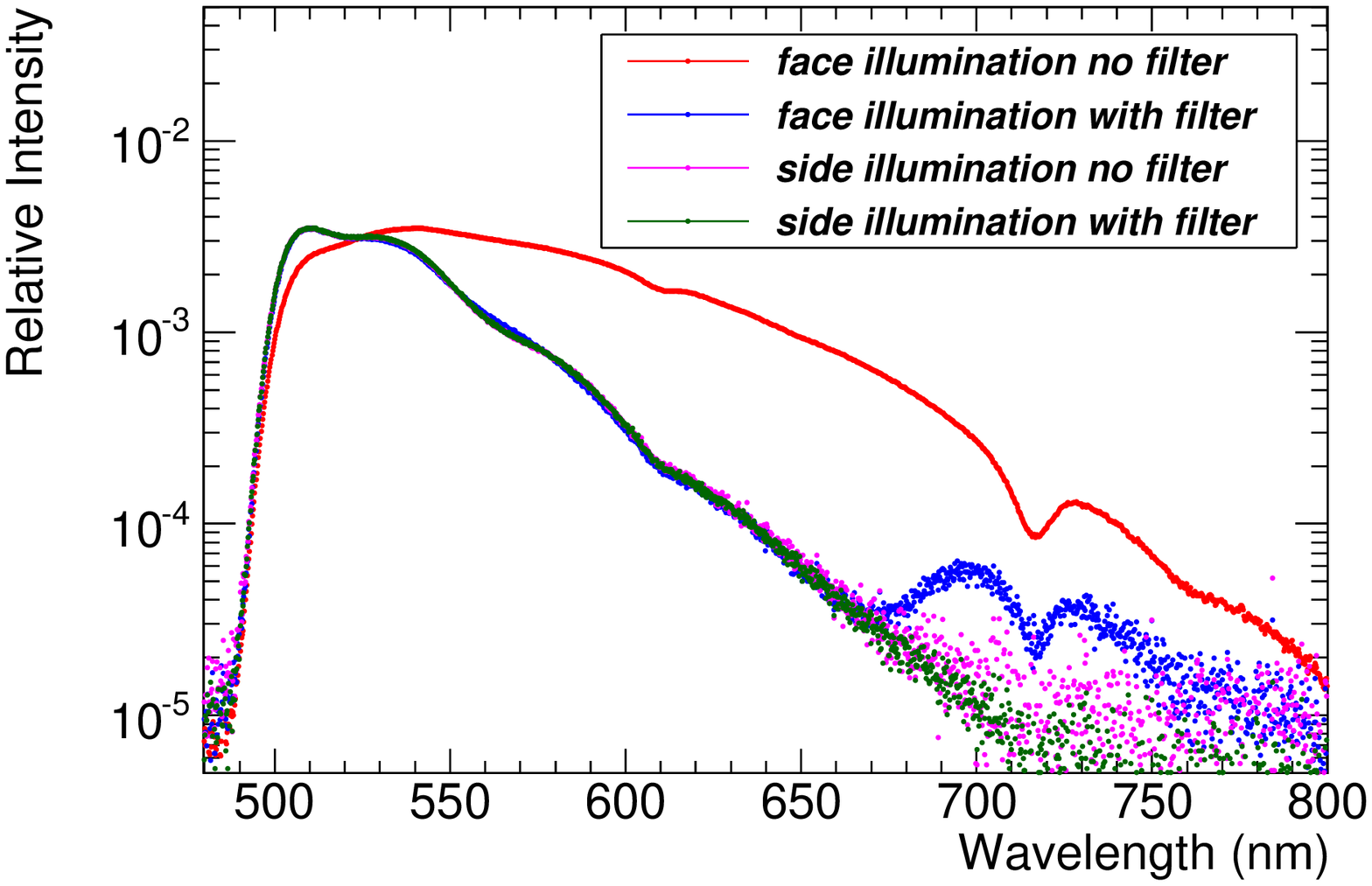} 
\caption{The effect of LED orientation. The fiber was illuminated from the side and from 
the face, with and without a low-pass ($<$ 500\,nm) filter. }
\label{fig:led_orientation}
\end{figure}

We studied the effect of illuminating the fiber from the side and from the end.  The light output spectrum depends heavily on the orientation of the LED.  Illuminating the fiber from the side results in transmission of long wavelengths and thus the observed spectrum is only from those photons that have undergone at least one wavelength shifting event.  Alternatively, illuminating the fiber on the face results in the propagation of all wavlengths.  A clear difference can be seen in Figure~\ref{fig:led_orientation} where we illuminate a 1\,m 0.7\,mm diameter 200 ppm fiber with a 395\,nm LED as seen in Figure~\ref{fig:led_spectra}.  This LED has a broad emission spectrum spanning most of the visible wavelength region.  To confirm that the long wavelength transmission contributes to the observed spectrum,  we placed a filter between the LED and the fiber then repeated the two illumination schemes.  The filter suppressed wavelengths larger than 500\,nm.  The results show that applying a filter to the face illumination scheme suppresses the longer wavelengths. The results also show that the output spectra from the fiber are the same for side illumination with and without a filter, and also for face illumination with a filter applied. 


\section{Additional modeling features}

\subsection{Dependence on diameter}

\begin{figure}[h]
\centering
\includegraphics[width=0.5\textwidth]{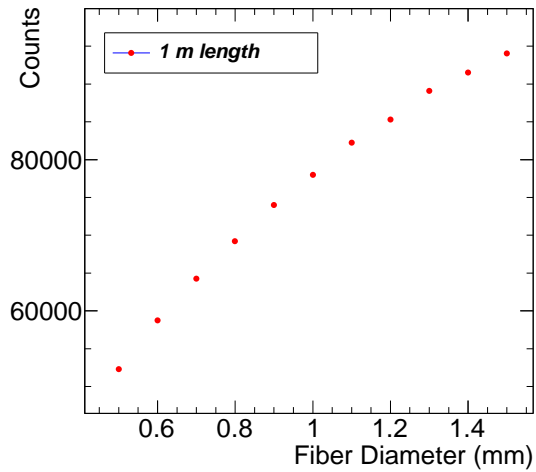} 
\caption{Collection efficiency versus fiber diameter for a 1\,m, 200\,ppm WLS fiber. }
\label{figure:fiber_diameters}
\end{figure}

We investigated the effect of fiber diameter on the spectral output at the end of the fiber and the attenuation rate.  The fiber diameter was varied from 0.5\,mm to 1.4\,mm in increments of 0.1\,mm.   Figure~\ref{figure:fiber_diameters} shows the results for the spectral output.  No wavelength dependence on diameter was observed.

\subsection{Dependence on concentration}

\begin{figure}[h]
\centering
\includegraphics[width=0.8\linewidth]{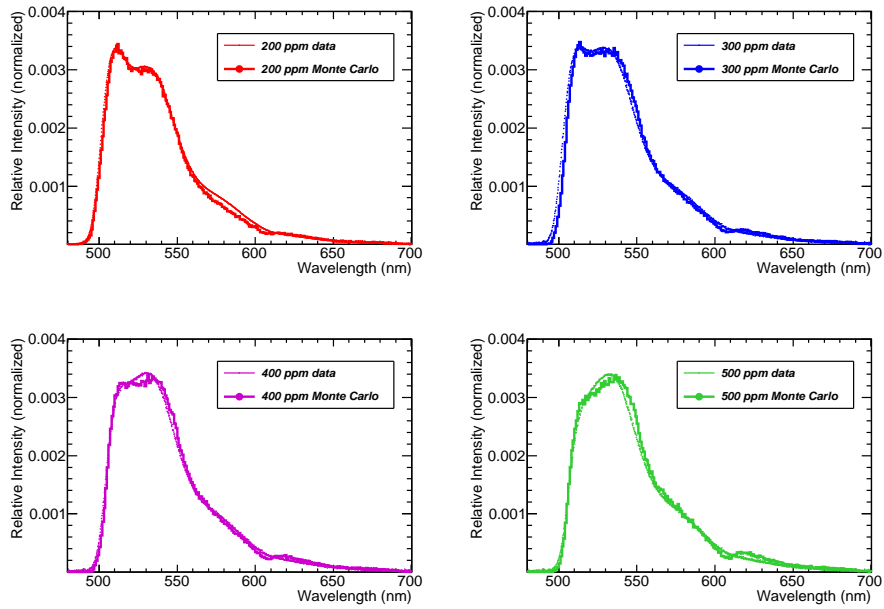}
\caption{Comparison of data and Monte Carlo for a 1\,m 0.7\,mm diameter WLS fiber for 
four concentrations (200, 300, 400, and 500\,ppm) of the K27 fluorophore.  }
\label{fig:300ppm_spectrum}
\end{figure}

Information on the complete spectral attenuation length as a function of concentration for the Y-11 fibers were unavailable.  Here, we used the model to predict the attenuation length of Y-11 at different concentrations of the K27 dopant. In the simulation, we varied the Y-11 absorption length by scaling the Y-11 absorption length in 5\% increments to determine if this quantity alone could reproduce measurements obtained using the Ocean Optics spectrophotometer.  We assume that the spectral profile is dependent only on the WLS-doped fiber absorption length.  Figure~\ref{fig:300ppm_spectrum} shows the Monte Carlo predicted and measured distributions of photons exiting the fiber for 200, 300, 400, and 500\,ppm concentrations.  We found that with this scaling, approximate agreement could be reached at X\% for 300\,ppm, Y\% for 400 ppm, and Z\% for 500\,ppm.

\subsection{Photon propagation time}

\begin{figure}[h]
\centering
\includegraphics[width=0.45\linewidth]{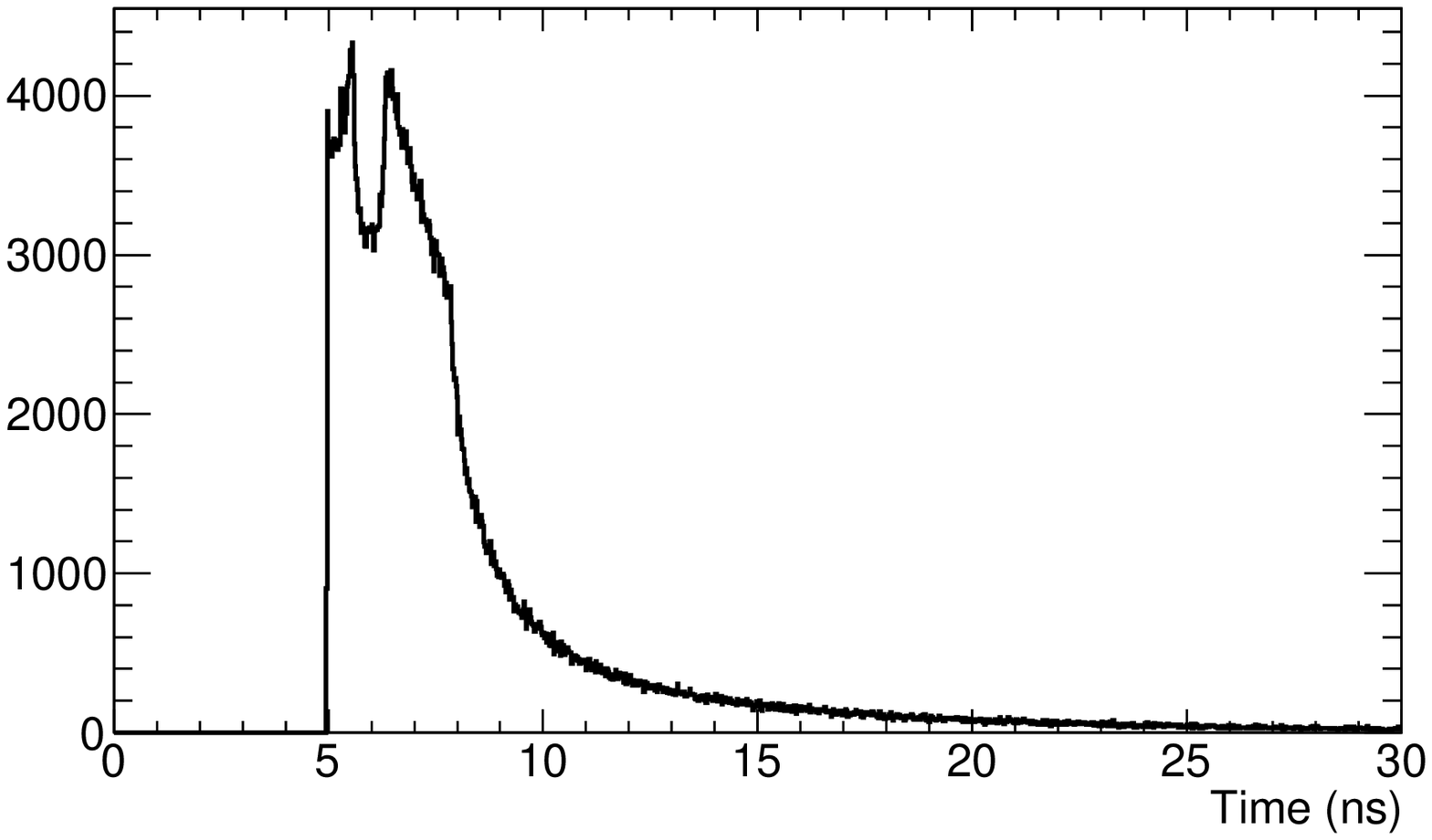}
\hskip0.25in
\includegraphics[width=0.45\linewidth]{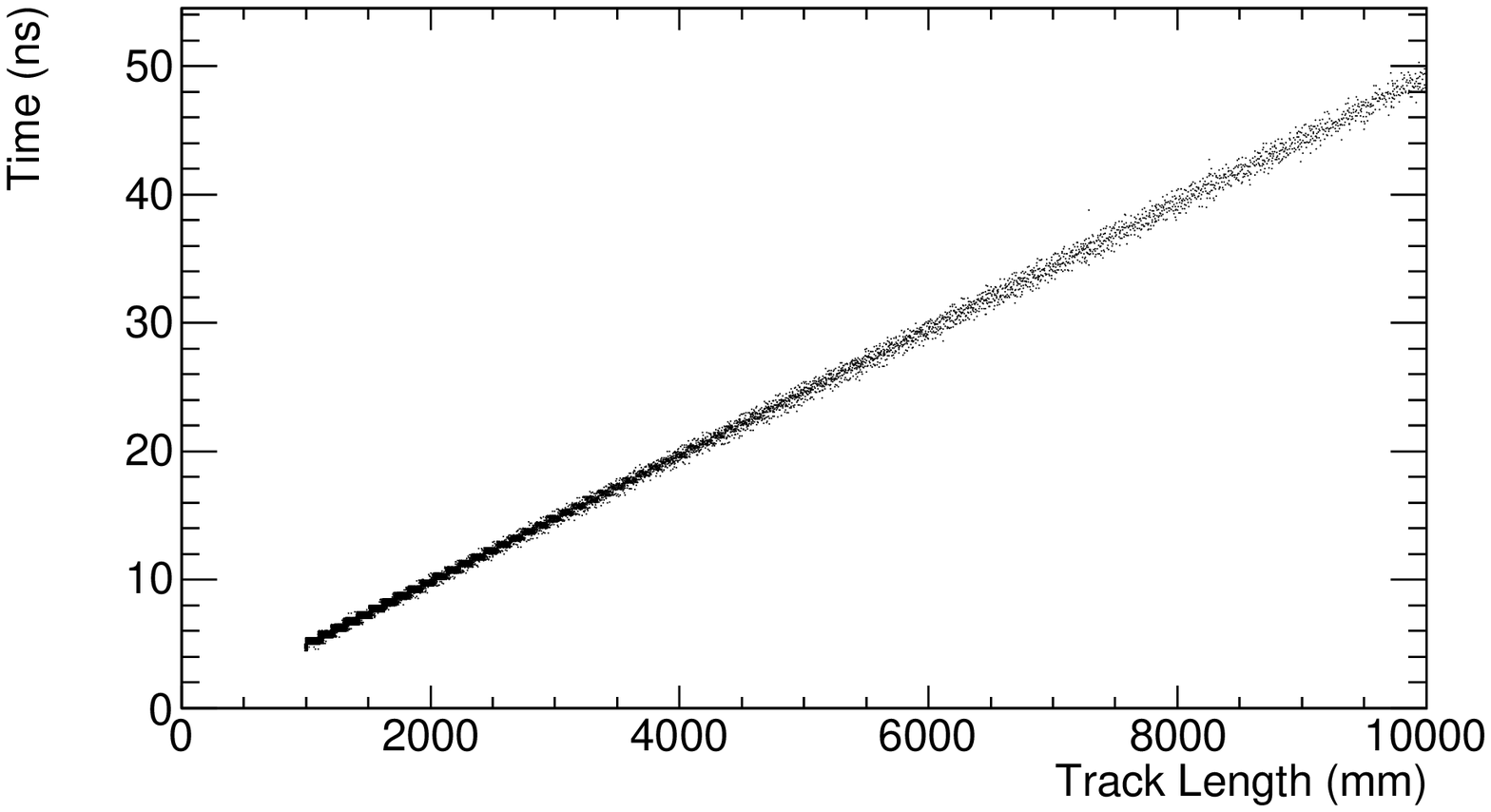}
\caption{Timing distribution for collected photons (left) and distance versus time (right) for a 1\,m 200\,ppm WLS fiber. }
\label{figure:timing_profiles}
\end{figure}

\begin{figure}[h]
\centering
\includegraphics[width=0.8\linewidth]{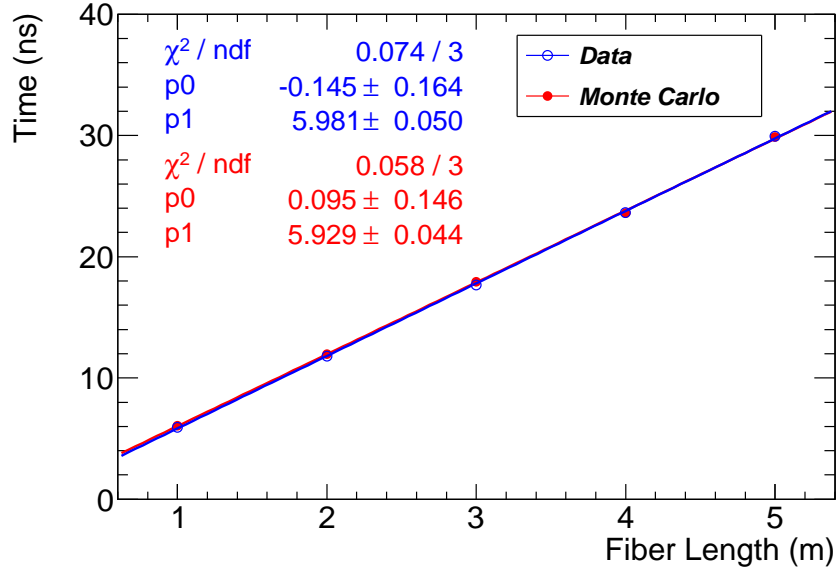}
\caption{The mean time for collected photons for several lengths of 200\,ppm WLS fiber.  
Data is shown in blue and Monte Carlo is shown in red.}
\label{figure:propagation_times}
\end{figure}

We assume a wavelength shifting time constant of 11.8\,ns~\cite{PlaDalmau:1994kx}.  Figure~\ref{figure:timing_profiles} shows the time distribution for collected photons and the total distance versus time.  We extract an effective refractive index of approximately 1.5 based on these results.  The jitter observed between 5 and 8\,ns is attributed to the various modes of propagation within the fiber.  Figure~\ref{figure:propagation_times} shows photon propagation times for several lengths of 200\,ppm WLS fiber.  The first peak is the result of photons which propagate and interact predominantly in the claddings while the second peak is the result of photons propagating predominantly in the core.

\subsection{Photon distributions}

\begin{figure}[h]
\begin{center}$
\begin{array}{cc}
\includegraphics[width=0.43\textwidth]{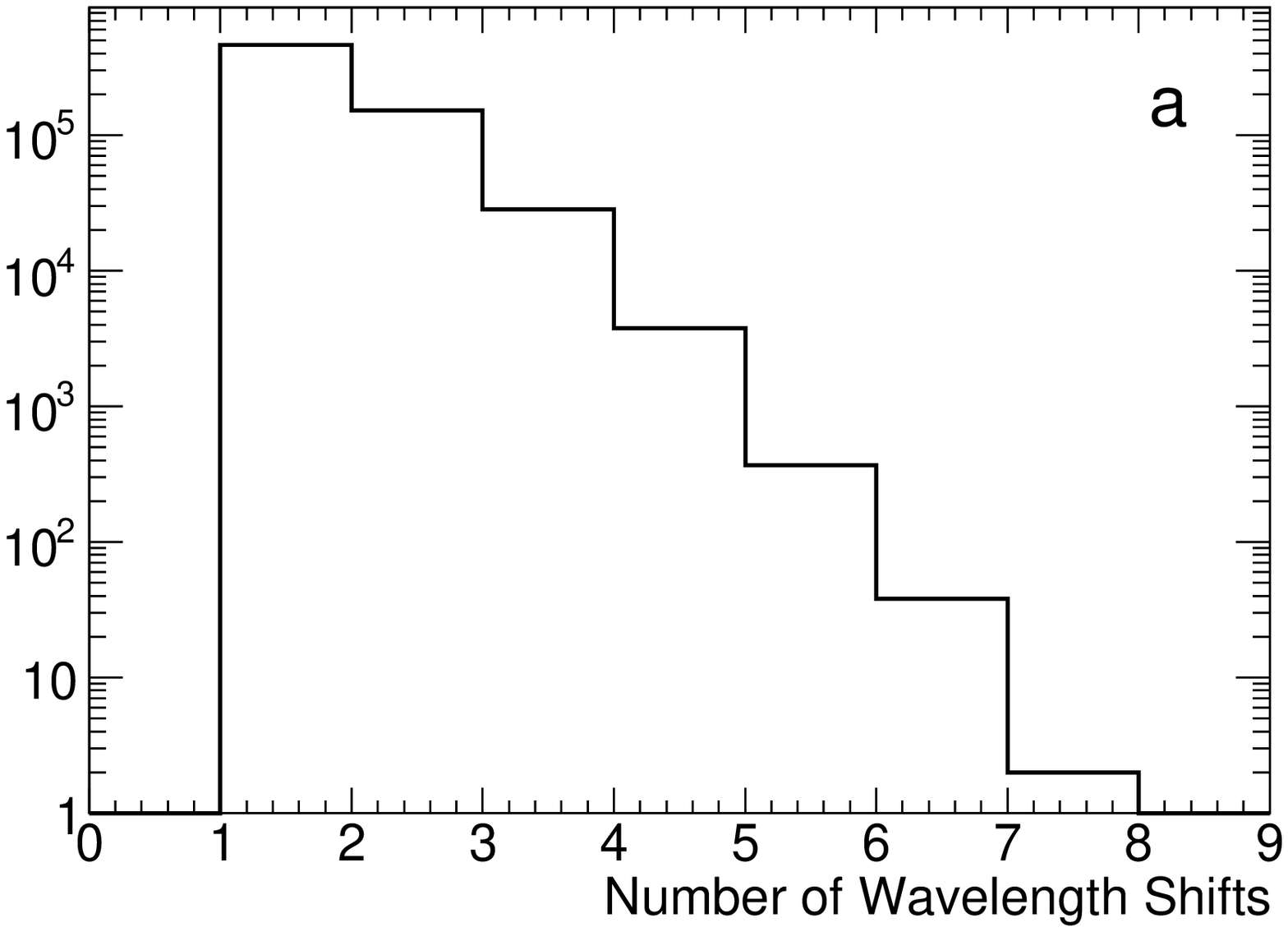} & 
\includegraphics[width=0.43\textwidth]{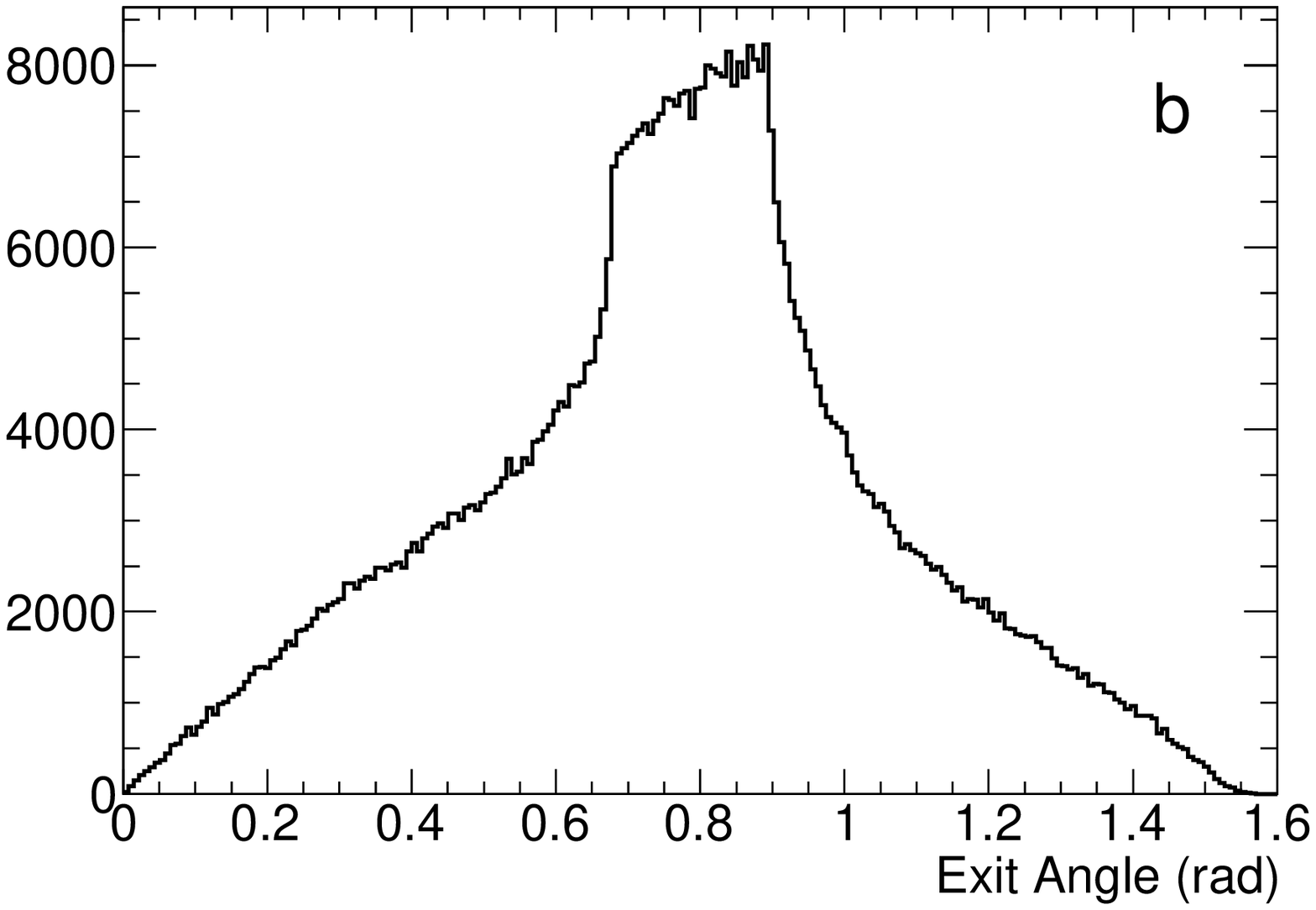} \\
\includegraphics[width=0.43\textwidth]{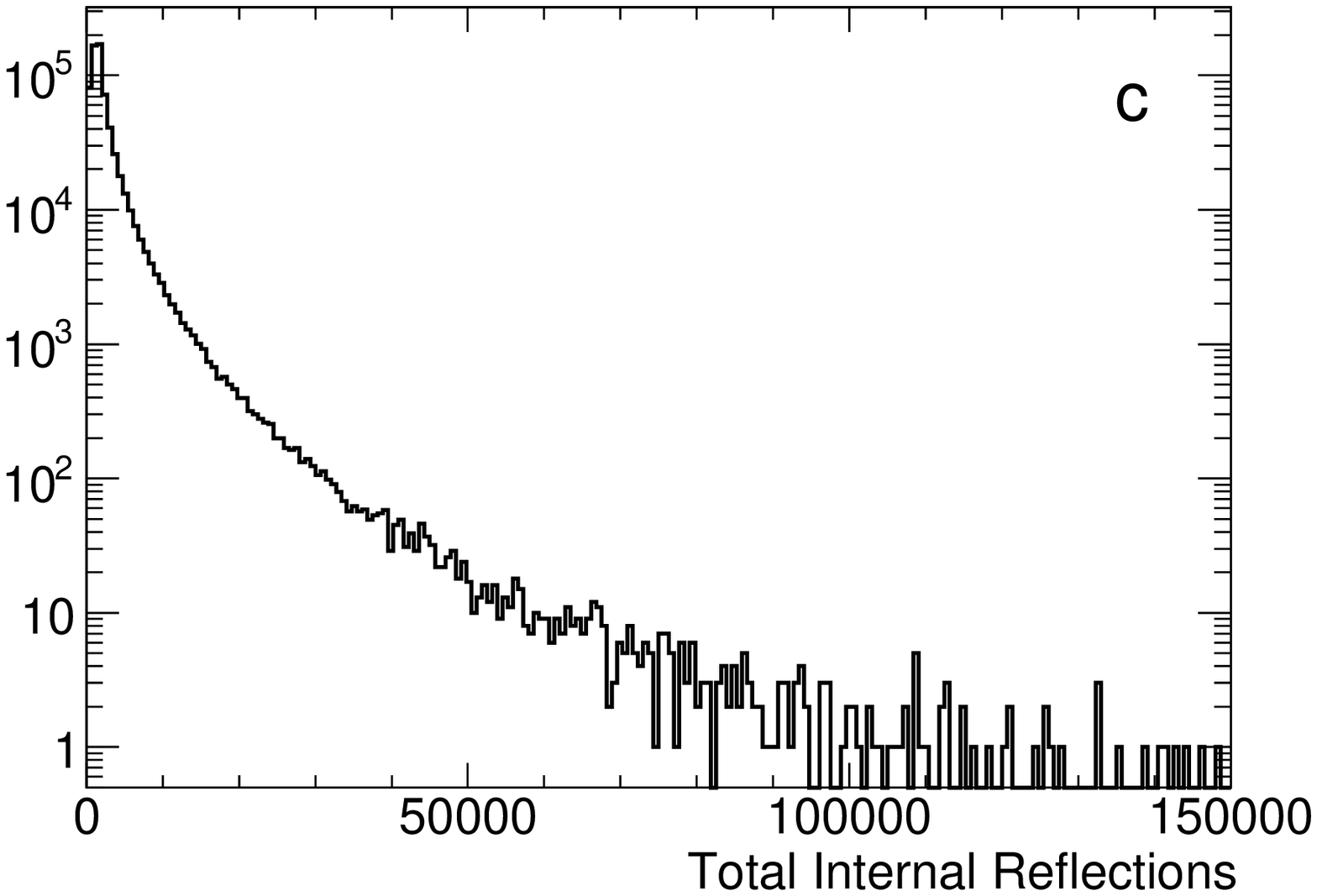} & 
\includegraphics[width=0.43\textwidth]{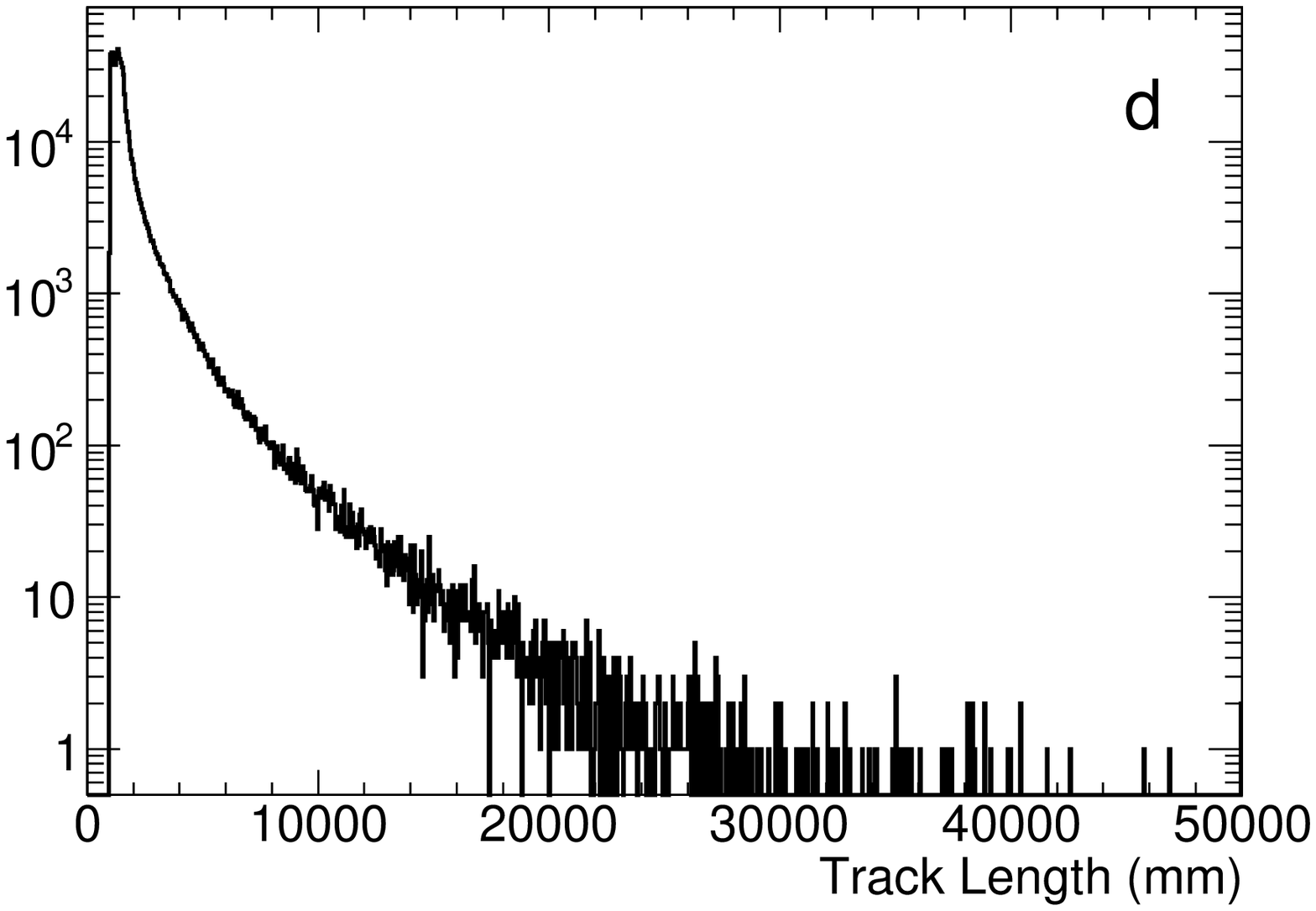} \\

\end{array}$
\end{center}
\caption{Several additional modeling results for collected photons from a 1 meter, 0.7\,mm diameter, 200\,ppm WLS fiber.  
a) The distributions of the number of wavelength shifts, 
b) the exit angle with respect to the principal fiber axis, c) the number of total internal reflections, and d) the total track length.}
\label{figure:wlscount_etc}
\end{figure}

Figure~\ref{figure:wlscount_etc} shows the distribution of the number of wavelength shifting interactions for collected photons.  All photons have undergone at least one wavelength shift while some have undergone more than five.  This figure also shows the distribution of the exit angle with respect to the principal fiber axis, the number of total internal reflections, and the total track length for collected photons.  


\begin{figure}[h]
\centering
\includegraphics[width=0.6\textwidth]{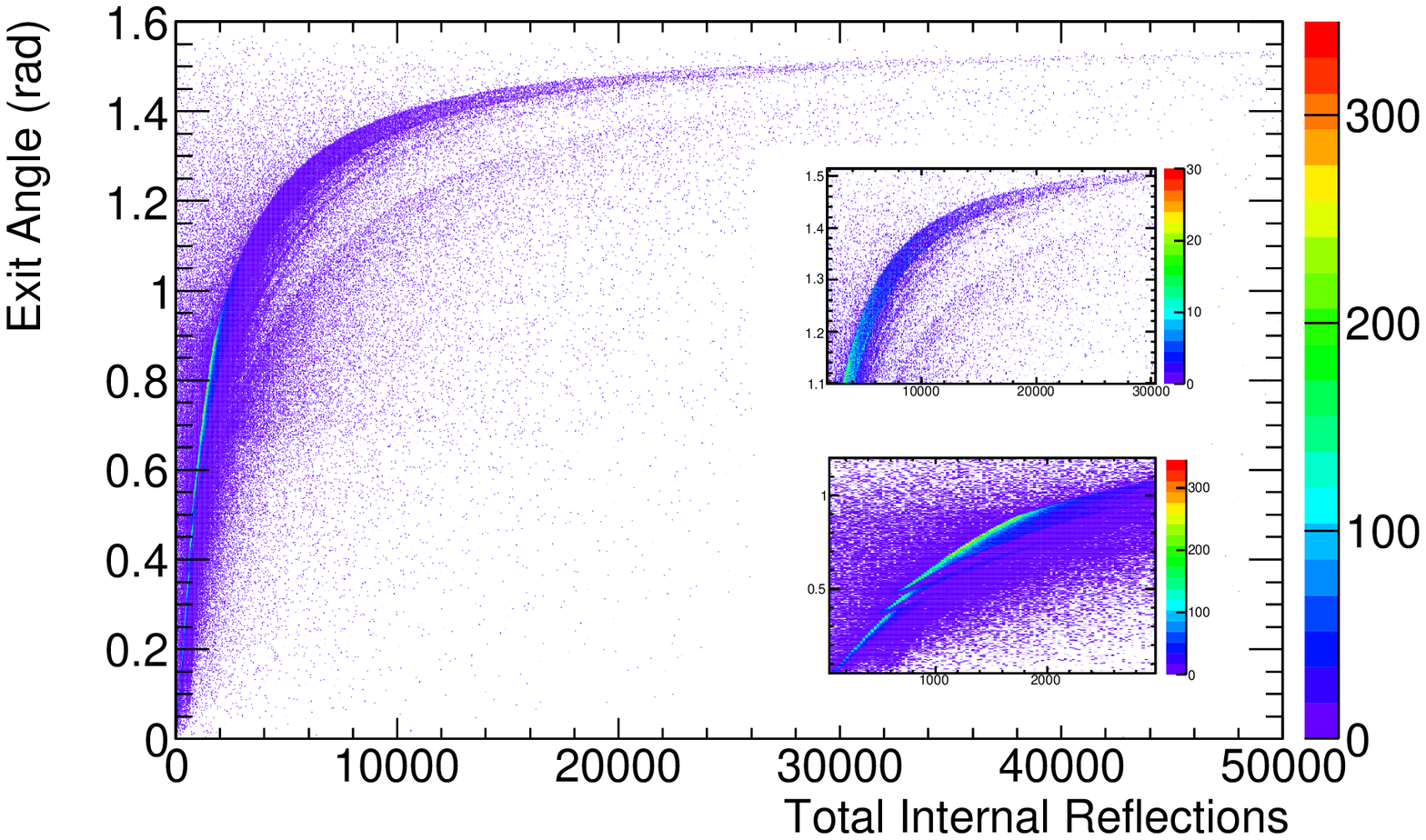}
\includegraphics[width=0.6\textwidth]{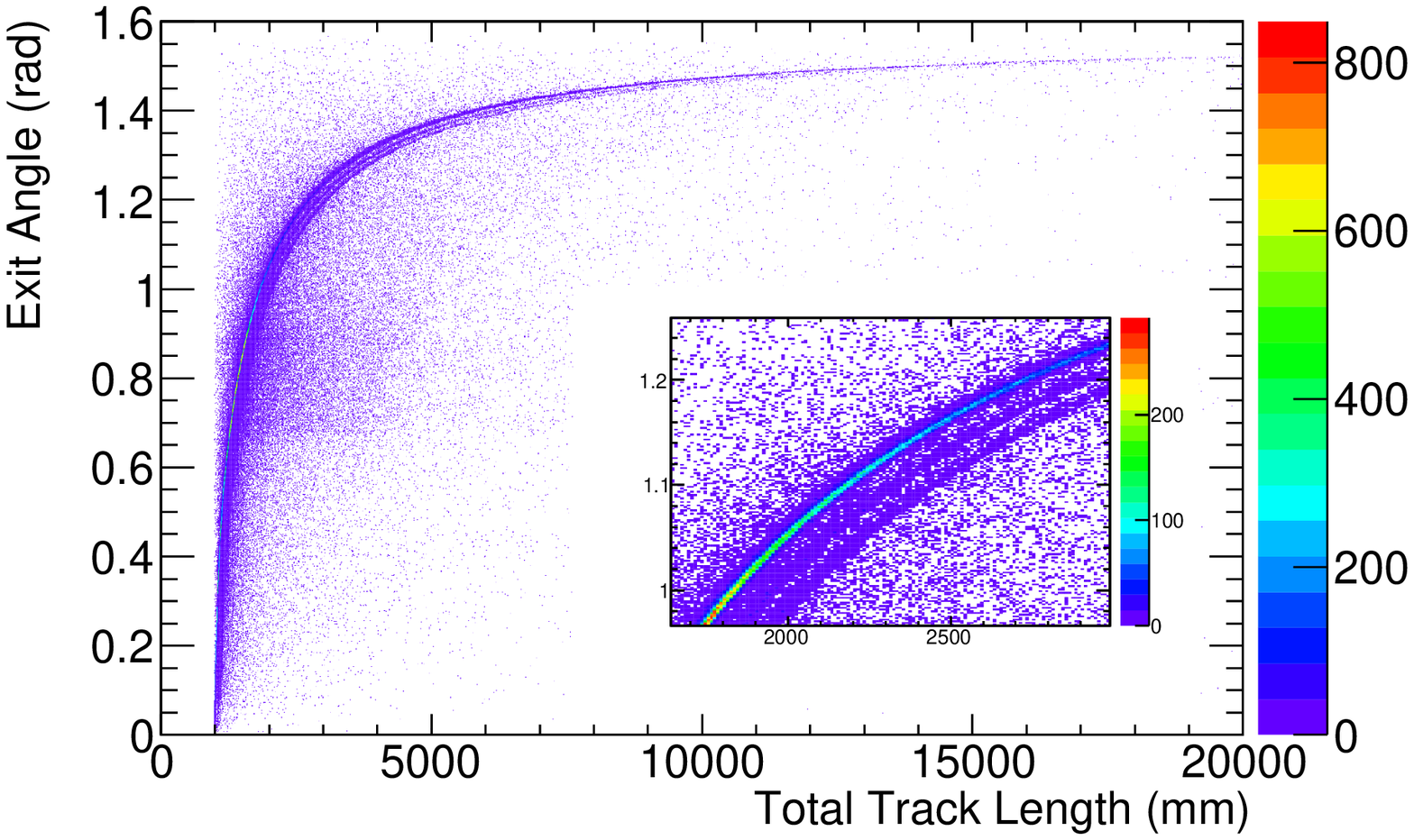}
\includegraphics[width=0.6\textwidth]{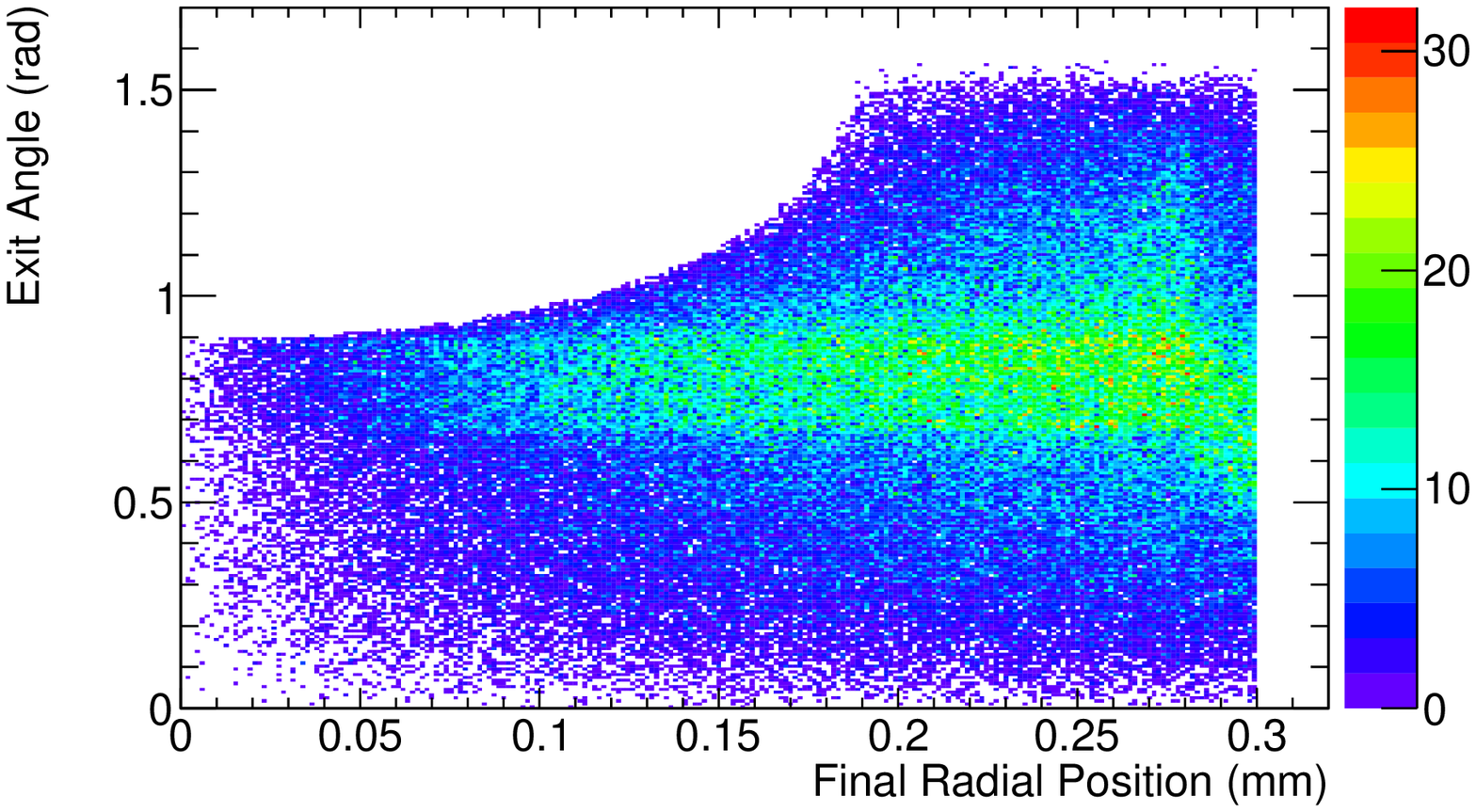}
\caption{The distributions of the number of wavelength shifts (top left), the exit angle with respect to the principal fiber axis (top right), the number of total internal reflections (bottom left), and the total track length (bottom right) for collected photons from a 1 meter, 0.7\,mm diameter, 200\,ppm WLS fiber. }
\label{figure:tir_vs_exitangle_etc}
\end{figure}

Figure~\ref{figure:tir_vs_exitangle_etc} shows the number of total internal reflections for collected photons as a function of their exit angle.   This figure also shows the distribution of track lengths for collected photons as a function of the exit angle, the final radial position of collected photons as a function of the final wavelength, and the distribution of the final radial position of collected photons as a function of the exit angle.

\section{Discussion and conclusion}

There are several limitations to the model.  We have considered an idealized fiber and have not introduced effects such as bubbles, dust, diameter variation over the fiber length, defects, or cracks.  These anomalies are macroscopically accounted for in the bulk attenuation length.  The effect of Rayleigh scattering has also been absorbed in the bulk attenuation length as explained previously~\cite{Bunge}.  From this, we expect idealized photon transport in the fiber which we speculate explains the dip in the fiber position profile that we model.  

We considered the possibility of the existence of a graded refractive index which can arise from non-uniform cooling during production, which may alter the propagation modes, potentially giving rise to an altered position profile.  We simulated this by modeling the core of the fiber as ten concentric cylindrical shells each with a different refractive index.  We scaled the wavelength dependent refractive index from 1.55 to 1.65 going radially outwards.  The results were nearly identical to the non-graded, constant case.  

We recognize that different manufacturers employ different production methods which can result in different optical characteristics.  We also recognize that different batches of the same production may give rise to different optical properties.  The simulations presented here are estimates based on Kuraray Y-11 data with a nominal K27 concentration of 200 ppm.  

We have not considered the effects of birefringence~\cite{Dugas} nor the effects of the K27 molecular orientation within the polymer~\cite{Llop}, both of which could potentially alter the primary emission spectrum and subsequent propagation.  It is known that the S-type fiber displays these characteristics~\cite{Kuraray_1}.  We found that illuminating the fibers from the end using these LEDs produced spectra that were somewhat different from those illuminated from the side.  This is due to the notion that the LEDs typically emit light over a broad spectrum which results in attenuation at short wavelengths and transmission at long wavelengths.  Therefore, we found that simulating the LEDs correctly was paramount. 
The model can be extended to any plastic based detector system employing fluorescent wavelength shifters, square fibers, and optical cuvettes provided one has the relevant input spectra including the doped and un-doped substrate and the corresponding bare emission spectrum (along with any refractive indices and reflection coefficients, if needed).  

In these studies, we introduce a general model of photon transport based on the GEANT4 framework that accounts for all of the processes mentioned above for clear and wavelength shifting plastic optical fiber.  We validate the model using measurements from Kuraray Y-11 wavelength shifting fibers of various lengths then use the model to predict the spectral output behavior from fibers with various concentrations, diameters, and bending radii.  We demonstrate that by including the spectral properties of all components and properly accounting for absorption, reemission, and fluorescent quantum yield of the chromophore, this model can accurately determine the spectral response, timing, and effective attenuation in wavelength shifting fibers.

Acknowledgments:
We thank Megan Creasey for assistance with the measurements. 
This research was supported in part by NSF grant PHY-0902235, and DOE grant DE-FG03-93ER40757.



\end{document}